\documentclass[PRD,twocolumn,showpacs,superscriptaddress,preprintnumbers,nofootinbib,amsmath,amssymb]{revtex4-1}

\usepackage[colorlinks=true,urlcolor=blue,linkcolor=blue,citecolor=blue]{hyperref}
\usepackage[normalem]{ulem}
\usepackage[capitalize]{cleveref}

\usepackage{comment}

\usepackage[T1]{fontenc} 

\usepackage{mathtools}

\usepackage{xcolor}
\usepackage{graphicx}
\usepackage{dcolumn}
\usepackage{bm}
\usepackage{multirow}

\usepackage{epstopdf}

\newcommand{\Amu}{\mathcal{A}^{\mu}}

\newcommand{\Ao}{\mathcal{A}^{0}}
\newcommand{\Ai}{\mathcal{A}^{i}}
\newcommand{\Aj}{\mathcal{A}^{j}}
\newcommand{\hc}{\text{h.c.}}
\newcommand{\Beq}{\begin{equation}\begin{aligned}}
\newcommand{\Eeq}{\end{aligned}\end{equation}}

\begin{document}

\title{Self-gravitating Vector Dark Matter}


\author{Peter Adshead}
\affiliation{Illinois Center for Advanced Studies of the Universe \& Department of Physics, University of Illinois at Urbana-Champaign, Urbana, IL 61801, USA.}
\author{Kaloian D. Lozanov}
\affiliation{Illinois Center for Advanced Studies of the Universe \& Department of Physics, University of Illinois at Urbana-Champaign, Urbana, IL 61801, USA.}



\begin{abstract}
We derive the non-relativistic limit of a massive vector field. We show that the Cartesian spatial components of the vector behave as three identical, non-interacting scalar fields. We find classes of spherical, cylindrical, and planar self-gravitating vector solitons in the Newtonian limit. The gravitational properties of the lowest-energy vector solitons---the gravitational potential and density field---depend only on the net mass of the soliton and the vector particle mass. In particular, these self-gravitating, ground-state vector solitons are independent of the distribution of energy across the vector field components, and are indistinguishable from their scalar-field counterparts. Fuzzy Vector Dark Matter models can therefore give rise to halo cores with identical observational properties to the ones in scalar Fuzzy Dark Matter models. We also provide novel hedgehog vector soliton solutions, which cannot be observed in scalar-field theories. The gravitational binding of the lowest-energy hedgehog halo is about three times weaker than the ground-state vector soliton. Finally, we show that no spherically symmetric solitons exist with a divergence-free vector field.
\end{abstract}


\maketitle

\section{Introduction}
The Lambda Cold Dark Matter ($\Lambda$CDM) model is extremely successful in describing the Universe on galaxy-cluster and cosmic scales ($\gg10$ kpc). The spectrum of density fluctuations inferred from the CMB at redshift $z\sim10^3$ is in agreement with the one observed today, $z\simeq0$, at the percent level \cite{Aghanim:2018eyx}. However, on galactic scales ($\lesssim10$ kpc), $\Lambda$CDM appears to be in tension with observations. N-body simulations predict density cusps in the cores of dark matter halos \cite{Navarro:1996gj}, whose signature is not seen in galaxy rotation curves \cite{Flores:1994gz,Salucci_2019}. Furthermore, the expected abundance of low-mass halos greatly exceeds the one inferred from observations of satellite galaxies \cite{Moore:1999nt,Klypin:1999uc,10.1111/j.1745-3933.2011.01074.x}. These shortcomings could be attributed to inaccurate modeling of baryonic physics \cite{Sawala:2015cdf,Read:2018fxs,DelPopolo:2018wrz}. However, no rigorous argument has yet been put forward that uses known physics to reconcile $\Lambda$CDM and galactic-scale observations.

One class of alternative solutions that address these shortcomings include modifications of the nature and properties of dark matter. Warm Dark Matter (WDM) scenarios assume a light dark-matter particle ($m\sim$ keV) whose thermal velocity dispersion suppresses small-scale structure \cite{Viel:2005qj}. The formation of density cusps and low-mass halos is prevented by free-streaming of the WDM out of potential wells \cite{Hirsch:2000ef,Bode:2000gq}. The small-scale distribution of dark matter can be also modified by introducing strong self-interactions \cite{Tulin:2017ara,Agrawal:2020lea}. 

Another popular class of  solutions to these small-scale problems is the Fuzzy Dark Matter (FDM) scenario. This scenario involves a free bosonic ultra-light ($m\sim10^{-22}-10^{-21}$ eV) non-relativistic dark-matter particle whose wave nature is manifested on astrophysical and galactic scales \cite{Hu:2000ke}. On scales comparable to the kpc de Broglie wavelength of the FDM particle, the formation of density cusps and low-mass halos is suppressed. On larger scales, FDM is indistinguishable from CDM. A compelling FDM candidate is an ultra-light axion. The right relic abundance of axion FDM with $m$ in the desired range can be achieved easily in natural High-Energy Physics models \cite{Hui:2016ltb}. 

Many aspects of the cosmology and astrophysics of  axion FDM  have received great attention in recent years. The occupation numbers within galaxy halos are so high (the de Broglie wavelength is much larger than the interparticle distance) that the state of the axion can be described as a classical non-relativistic scalar-field condensate. The phenomenology of the self-gravitating condensate has been widely studied in the Newtonian limit of gravity \cite{PhysRev.187.1767,SPIEGEL1980236,PhysRevD.42.384,Widrow:1993qq,Peebles:2000yy,Boehmer:2007um,Woo:2008nn,Chavanis_2011,Chavanis_2011b,Guth:2014hsa,Schive:2014dra,Marsh:2015xka,terazawa2015exotic,Schwabe:2016rze,Mocz:2017wlg,Li:2018kyk,Zhang:2018ghp,Lancaster:2019mde,Kendall:2019fep,Musoke:2019ima,Niemeyer:2019gab,Poddar_2020,Schwabe:2020eac,Annulli_2020,Annulli_2020b,Eggemeier:2020zeg,May:2021wwp}. This self-gravitating condensate leads to a number of astrophysical signatures (for a recent review, see Ref. \cite{Ferreira:2020fam}), most of which are a consequence of the fact that the cores of FDM halos are the ground state soliton solution of the Schr\"{o}dinger-Poisson system. 

FDM could also be realized by higher-spin bosons \cite{Nelson:2011sf,Arias:2012az,Dimopoulos:2006ms,Poddar_2021}. In this work we consider an ultra-light spin-1 boson, minimally coupled to gravity; a Vector Dark Matter (VDM) particle. We derive its non-relativistic\footnote{For a related work in Minkowski spacetime of lower dimensions see Ref. \cite{Bergshoeff:2018tjg}.} Newtonian limit and show that the Fuzzy VDM condensate is a collection of three scalar-field condensates, interacting only gravitationally. We study soliton solutions of the resulting equations, and demonstrate that at least one VDM soliton has an identical distribution of mass (and therefore gravitational potential) to its axion counterpart. For this class of solutions,  no distinction can be made between axion FDM halos and Fuzzy VDM halos on the basis of their gravitational properties in the Newtonian limit.

Throughout this work we remain agnostic about the High-Energy Physics embedding of VDM. We study the massive vector field at the phenomenological level, without specifying the origin of  its mass  (Higgs or Stueckelberg)  or how it is produced in the primordial universe (see, for example, Refs. \cite{Graham:2015rva,Agrawal:2018vin,Dror:2018pdh,Bastero-Gil:2018uel,Co:2018lka,Ema:2019yrd,Kolb:2020fwh,Ahmed:2020fhc}).

The rest of the paper is organized as follows. In Section \ref{sec:Model} we outline the relativistic theory of a massive vector field and derive its equations of motion in a perturbed Friedmann-Robertson-Walker (FRW) background. In Section \ref{sec:NRVDN} we study the non-relativistic regime of the theory in the Newtonian gravity limit. The classes of Fuzzy VDM solitons are described in Section \ref{sec:VDMHalos}. We present our concluding remarks in Section \ref{sec:Conclusions}. In Appendix \ref{App:Soliton} we provide details for the numerical procedure used to find vector soliton solutions. We show that Fuzzy VDM can be considered as a three-component superfluid in Appendix \ref{App:SFVDM}.

Throughout this work we use the Einstein summation convention for repeated indices. Repeated upper and lower Greek indices (space-time indices) are summed over $\mu \in \{0,1,2,3\}$, while repeated latin indices (spatial indices) of any level are summed over $i\in \{1,2,3\}$. We do not use the summation convention for non-vector quantities which have an index inside a bracket, for example $I^{(j)}$. We adopt natural units in which $\hbar=c=1$ and the reduced Planck mass $m_{\rm Pl}=1/\sqrt{8\pi G}=2.435\times10^{18}$ GeV.

\section{The model}
\label{sec:Model}

Consider a massive vector field (sometimes referred to as a Proca field) minimally coupled to gravity with an action
\Beq
\label{eq:Action}
S=\int d^4x\sqrt{-g}\left[-\frac{m_{\rm Pl}^2R}{2}-\frac{1}{4}F^{\mu\nu}F_{\mu\nu}+\frac{1}{2}m^2A^{\mu}A_{\mu}\right]\,,
\Eeq
where the Ricci scalar, $R=R^\alpha{}_\alpha$, is obtained from the trace of the Ricci tensor.
\Beq
R_{\alpha\beta}=\partial_\delta \Gamma^{\delta}{}_{\alpha\beta}-\partial_\beta \Gamma^{\delta}{}_{\alpha\delta}+\Gamma^\delta{}_{\alpha\beta}\Gamma^{\sigma}{}_{\delta\sigma}-\Gamma^\sigma{}_{\alpha\delta}\Gamma^\delta{}_{\beta\sigma}\,,
\Eeq
and the field-strength tensor for the vector field $A_\mu$ is
\Beq
F_{\mu\nu}=\nabla_{\mu}A_{\nu}-\nabla_{\nu}A_{\mu}=\partial_{\mu}A_{\nu}-\partial_{\nu}A_{\mu}\,.
\Eeq
The Euler-Lagrange equations that follow from variation of the action in eq.\ \eqref{eq:Action} yield the Einstein equations
\Beq
\label{eq:Einstein}
G_{\mu\nu}=R_{\mu\nu}-\frac{g_{\mu\nu}}{2}R=\frac{1}{m_{\rm Pl}^2}T_{\mu\nu}\,,
\Eeq
where
\Beq
\label{eq:EMtensor}
T^\mu{}_\nu=m^2A^{\mu} A_{\nu}&-F^{\mu\alpha}F_{\nu\alpha}\\
&+\delta^\mu_\nu\left[-\frac{m^2}{2}A_\alpha A^\alpha+\frac{1}{4}F^{\alpha\beta}F_{\alpha\beta}\right]\,,
\Eeq
and the Maxwell equations
\Beq
\label{eq:Maxwell}
\nabla_{\mu}F^{\mu\nu}=J^\nu\,,
\Eeq
where
\Beq
J^\nu=-m^2A^\nu\,.
\Eeq
Despite the fact that the action in eq.\ \eqref{eq:Action} does not possess a $U(1)$ gauge symmetry (due to the mass term), $A^{\nu}$ respects an equation identical to the Lorenz gauge condition in free theories
\Beq
\label{eq:Lorenz}
\nabla_{\nu}A^{\nu}=0\,.
\Eeq
This equation follows from contracting eq.\ \eqref{eq:Maxwell} with $\nabla_{\nu}$.   In what follows, we refer to eq.\ \eqref{eq:Lorenz} as the Lorenz constraint. It is a manifestation of the fact that among the four components of $A^{\mu}$ there are only three degrees of freedom. Colloquially, $A^0$ is the non-dynamical (or auxiliary) field, since it does not have a kinetic term in the action in eq.\ \eqref{eq:Action}, whereas the $A^i$ capture the two transverse and one longitudinal physical degrees of freedom (the longitudinal one being present due to the non-vanishing mass of the vector field).

\subsection{The Weak Gravity limit}
The perturbed metric in a flat FRW background takes the following form
\Beq
ds^2&=a^2(\tau)\Big[(1+2\Phi)d\tau^2+2(\partial_i C+V_i)dx^id\tau\\
&-\left[(1-2\Psi)\delta_{ij}-\partial_i\partial_j U - \partial_{[i}K_{j]}-h_{ij}\right]dx^idx^j\Big]\,,
\Eeq
where $\Phi$, $\Psi$, $C$, and $U$ are scalar perturbations, $V_i$ and $K_i$ are divergence-free vector perturbations ($\partial_i V_i = \partial_i K_i = 0$) and $h_{ij}$ is the transverse-traceless spatial metric perturbation ($\partial_i h_{ij} = \partial_i h_{ji} = 0$, and $h_{ii} = 0$). The two dynamical degrees of freedom of the linearized metric are the transverse traceless, or gravitational wave modes, $h_{ij}$.  We  ignore these in what follows, since they do not affect the vector field dynamics in the linear regime.  We retain only the scalar and vector perturbations which, in principle, can influence the dynamics of the vector field in the linear regime (when perturbing around an $\bar{A}^\mu$ background). 

To proceed, we fix the gauge freedom coming from the invariance of the action in eq.\ \eqref{eq:Action} under diffeomorphisms
\Beq
x^{\mu}\rightarrow x^{\mu}+\xi^\mu\,,
\Eeq
where $\xi^\mu=[\xi^0,\partial_i\xi^{(L)}+\xi_i^{(T)}]^T$. This invariance allows us to fix two scalars and one transverse vector. For the scalars, we choose to work in the Newtonian gauge
\Beq
C=0\,,\qquad\quad \text{and}\qquad\qquad U=0\,,
\Eeq
and we fix the transverse vector degree of freedom by setting
\Beq
K_i=0\,.
\Eeq
The remaining scalar and vector degrees of freedom ($\Phi$, $\Psi$, and $V_i$) do not represent true dynamical degrees of freedom (their action has no kinetic term), but are constrained variables whose evolution is determined by the dynamical variables, $A_\mu$ and $h_{ij}$.

The inverse and determinant of the metric (to linear order in perturbations) are then
\Beq
g^{\mu\nu}=\frac{1}{a^{2}(\tau)}\begin{pmatrix*}[c]1-2\Phi & V_1& V_2& V_3\\ V_1& -1-2\Psi & 0 & 0\\ V_2& 0 & -1-2\Psi & 0\\ V_3& 0 & 0& -1-2\Psi \end{pmatrix*}\,,
\Eeq
and
\Beq
\sqrt{-g}=a^4(\tau)\Big[1+(\Phi-3\Psi)\Big]\,,
\Eeq
respectively. We do not set $\Phi$ to be equal to $\Psi$, since in principle the vector field could have a sizable anisotropic stress, comparable to its isotropic pressure.

After plugging in the Lorenz constraint, eq.\ \eqref{eq:Lorenz}, into the Maxwell equation, eq.\ \eqref{eq:Maxwell}, one arrives at
\Beq
\label{eq:MaxwellLorenz}
\Box A^{\nu}+m^2 A^\nu+R^{\nu}_\mu A^\mu=0\,,
\Eeq
which when written out explicitly becomes
\Beq
\label{eq:MLEOM}
\frac{1}{\sqrt{-g}}\partial_{\mu}&\left(\sqrt{-g}g^{\mu\alpha}\partial_\alpha A^\nu\right)+m^2 A^\nu\\
&+A^\mu g^{\nu\alpha}\partial_\alpha\left[\frac{\partial_\mu \sqrt{-g}}{\sqrt{-g}}\right]-\partial_\alpha A^\mu\partial_\mu g^{\nu\alpha}=0\,.
\Eeq
To leading order in metric perturbations eq.\ \eqref{eq:MLEOM} reduces to
\Beq
\label{eq:EOM}
\text{for   }\nu=&0:\\
\partial_0^2 A^0&+2\mathcal{H}\partial_0 A^0-\left(1+2(\Phi+\Psi)\right)\Delta A^0\\
&-\partial_0\left(\Phi+3\Psi\right)\partial_0 A^0+2\mathcal{H}V_i\partial_i A^0\\
&+2V_i\partial_i\partial_0A^0-\partial_i(\Phi-\Psi)\partial_iA^0\\
&+(\partial_0 V_i)\partial_i A^0+(1+2\Phi)a^2m^2 A^0\\
&+ A^\mu\partial_\mu\partial_0(\Phi-3\Psi)+4A^0\partial_0\mathcal{H}\\
&+2(\partial_0A^\mu)\partial_\mu \Phi-(\partial_i A^\mu)\partial_\mu V_i\\
&+2\mathcal{H}\partial_0 A^0+2\mathcal{H}V_i\partial_i A^0=0\,,\\
\\
\text{for   }\nu=&j:\\
\partial_0^2 A^j&+2\mathcal{H}\partial_0 A^j-\left(1+2(\Phi+\Psi)\right)\Delta A^j\\
&-\partial_0\left(\Phi+3\Psi\right)\partial_0 A^j+2\mathcal{H}V_i\partial_i A^j\\
&+2V_i\partial_i\partial_0A^j-\partial_i(\Phi-\Psi)\partial_iA^j\\
&+(\partial_0 V_i)\partial_i A^j+(1+2\Phi)a^2m^2 A^j\\
&+\Big[A^\mu\partial_\mu\partial_i(\Phi-3\Psi)-2(\partial_iA^\mu)\partial_\mu \Psi\\
&+(\partial_0 A^\mu)\partial_\mu V_i+2\mathcal{H}(1+2(\Phi+\Psi))\partial_i A^0\\
&-2\mathcal{H}V_i\partial_0 A^0\Big]\delta^{ij}=0\,,
\Eeq
where $\mathcal{H}=\partial_0a/a$ and $\Delta=\partial_i\partial_j\delta^{ij}$. The first two terms in eq.\ \eqref{eq:MLEOM} and the first four lines in each of the equations in eq.\ \eqref{eq:EOM} are identical to the case of four copies of a massive scalar field in a spacetime with scalar and vector metric perturbations. However, the fact that $A^{\nu}$ is a spacetime vector field with a covariant derivative which mixes its components gives rise to the last two terms in eq.\ \eqref{eq:MLEOM} and the last three lines in each of the equations in eq.\ \eqref{eq:EOM}, which are not present in theories with scalar matter fields.

Another equation which is not present in scalar matter theories is the Lorenz constraint, eq.\ \eqref{eq:Lorenz}, which to leading order in metric perturbations yields
\Beq
\label{eq:LorenzLinear}
\partial_0 A^0+4\mathcal{H}A^0+\partial_i A^i &+A^0\partial_0\left(\Phi-3\Psi\right)\\
&+A^j\partial_j(\Phi-3\Psi)=0\,.
\Eeq

The remaining set of equations governing the evolution of the metric perturbations comes from the Einstein equation, eq.\ \eqref{eq:Einstein}. Its $00$, $ij$ and  transverse $0i$ components yield, respectively,
\Beq
&\Delta \Psi-3\mathcal{H}(\partial_0\Psi+\mathcal{H}\Phi)=\frac{a^2}{2m_{\rm Pl}^2}\delta T^0{}_0\,,\\
&\frac{2}{3}\Delta^2(\Phi-\Psi)=\frac{a^2}{m_{\rm Pl}^2}\left[\delta^{il}\delta^{jk}\partial_i\partial_j -\frac{\delta^{kl}\Delta}{3}\right]\delta T^k{}_l\,,\\
&\Delta V_i=\frac{2a^2}{m_{\rm Pl}^2}\delta T^{0}{}_i^{(T)}\,,
\Eeq
where on the right-hand side in the first and third lines we remove the spatial mean of the energy momentum tensor components  ($\delta T^0_{\,\,\,0}\equiv T^0_{\,\,\,0}-\overline{T^0_{\,\,\,0}}$, etc). Note also that $\partial_i T^{0}{}_i^{(T)}=0$. As expected, all scalar and vector metric perturbations are governed by constraint equations, since they do not represent true dynamical degrees of freedom.

\section{Non-relativistic VDM}
\label{sec:NRVDN}

In the non-relativistic limit we redefine the real vector field $A^\mu$ in terms of a complex field\footnote{Note that $\Amu$ is a four component complex field, but not a spacetime vector field. However, the spatial components, $\Ai$, constitute a complex 3-vector field.} $\Amu$ and its rest energy contribution\footnote{Without loss of generality, constant phase shifts between the individual components are absorbed into $\Amu$.}
\Beq
\label{eq:AmuNR}
A^\mu=\frac{1}{\sqrt{2m}}\left[\Amu e^{-im\int_\tau a(\tau') d\tau'}+\hc\right]\,.
\Eeq
We assume that the wave nature of the vector particles is manifest on co-moving sub-horizon scales, $2\pi/k$, for which $a^2m^2\gg k^2 \gg\mathcal{H}^2\sim|\partial_0\mathcal{H}|$. We further assume that all quantities ($\Amu$, $\mathcal{H}$, $\Phi$, $\Psi$ and $V_i$) vary slowly. That is, they do not change appreciably over a timescale of order the oscillation period $2\pi/m$. 

We proceed by time-averaging the product of $\exp({im\int_\tau a(\tau') d\tau'})$ and the Lorenz constraint, eq.\ \eqref{eq:LorenzLinear}. To leading order in metric perturbations and spacetime derivatives eq.\ \eqref{eq:LorenzLinear} reduces to
\Beq
\label{eq:LorenzFinal}
{iam\Ao=\partial_j\Aj}\,.
\Eeq

Then to leading order in spacetime derivatives  the time-averaged energy momentum reads
\Beq
\label{eq:EMNonR}
\langle T^\mu{}_\nu\rangle=\delta^\mu_0\delta^0_\nu m{\Aj}^*\!\Aj\,,
\Eeq
implying $V_i=0$, $\Phi=\Psi$ and\footnote{Note that $T_{ij} = 0$ at this order, and thus gravitational waves are not sourced at this order in the expansion in $k/(am)$.}
\Beq
\label{eq:PoissonFinal}
{\frac{\Delta}{a^2} \Phi=\frac{m{\Aj}^*\!\Aj}{2m_{\rm Pl}^2}}\,.
\Eeq
After applying the same considerations to the equations of motion, eq.\ \eqref{eq:EOM}, we find that they reduce to a set of Schr\"{o}dinger-type equations\footnote{We define cosmic time as $dt=ad\tau$ and the Hubble rate as $H=\dot{a}(t)/a(t)$, where an overdot denotes a derivative with respect to cosmic time, $t$.}
\Beq
\label{eq:EOMSchr0Final}
{\left[i\left(\partial_t+\frac{5}{2}H\right)+\frac{\Delta}{2a^2m}-\Phi m\right]\Ao=-i\Aj\frac{\partial_j}{a} \Phi}\,,\\
\Eeq
and
\Beq
\label{eq:EOMSchrjFinal}
{\left[i\left(\partial_t+\frac{3}{2}H\right)+\frac{\Delta}{2a^2m}-\Phi m\right]\Aj=0}\,.
\Eeq
Eq.\ \eqref{eq:EOMSchrjFinal} implies that at the homogeneous level, the energy density of each vector field component redshifts like matter, $m|\Aj|^2\propto a^{-3}$ .

Our Schr\"{o}dinger equations are consistent with the Lorenz constraint. If we take the divergence of eq.\ \eqref{eq:EOMSchrjFinal} and use eq.\ \eqref{eq:LorenzFinal} we arrive at eq.\ \eqref{eq:EOMSchr0Final}. 

One can now solve the Schr\"{o}dinger-Poisson system, eqs. \eqref{eq:PoissonFinal} and \eqref{eq:EOMSchrjFinal}, for $\Aj$ and $\Phi$, while making use of the Lorenz constraint, eq.\ \eqref{eq:LorenzFinal}, only to set up the initial conditions for $\partial_j\Aj$.\footnote{Besides when setting up the initial conditions for the longitudinal part of $\Aj$, there is no need to consider $\Ao$. It does not play a role in the evolution of the metric perturbations and $\Aj$. At any later moment we can consistently determine the value of $\Ao$ from the Lorenz constraint, eq.\ \eqref{eq:LorenzFinal}, and its time derivative from its Schr\"{o}dinger equation, eq.\ \eqref{eq:EOMSchr0Final}.} In fact, eqs. \eqref{eq:PoissonFinal} and \eqref{eq:EOMSchrjFinal}, can be derived from the Lagrangian density
\Beq
\label{eq:LagrangeSpatial}
\mathcal{L}={\Aj}^{*}\Big[i\partial_t&+\frac{3}{2}H-m\Phi\Big]\Aj\\
&-\frac{1}{2ma^2}|\nabla\Aj|^2 + \frac{m_{\rm pl}^2}{ a^2} \Phi \Delta \Phi\,.
\Eeq 
This Lagrangian density is invariant under a global $SU(3)$ transformation of the $\Aj$ components, 
\Beq
\label{eq:SU3Transf}
\Aj({\bf x},t)\rightarrow U^{ji}\Ai({\bf x},t)\,,\quad U\in SU(3)\,.
\Eeq
The gravitational potential is also $SU(3)$-invariant, $\Phi\rightarrow\Phi$. The emergent $SU(3)$ symmetry in the spatial sector in the non-relativistic limit of the theory is a consequence of the fact that we work with the complex 3-vector field, $\Aj$. Recall that $SU(2)$ is a subgroup of $SU(3)$, and is the double cover of $SO(3)$, which is the global symmetry of the non-relativistic limit of the (physical) real 3-vector field, $A^i$. The remainder of the $SU(3)$ is just a redundancy due to the fact that we are expressing the real theory in terms of a complex representation, see eq.\ \eqref{eq:AmuNR}. Nevertheless, this internal $SU(3)$ symmetry proves useful when we look for soliton solutions of the vector Schr\"{o}dinger-Poisson system, eqs. \eqref{eq:PoissonFinal} and \eqref{eq:EOMSchrjFinal}. 

Before we consider the soliton solutions, we note that the right-hand side of the Lorenz constraint, eq.\ \eqref{eq:LorenzFinal}, is not invariant under any of the $SU(3)$ transformations in eq.\ \eqref{eq:SU3Transf}. Hence, $\Ao$ is not an $SU(3)$ singlet. However, $\Ao$ transforms as a scalar under global coordinate rotations, $R$, belonging to the $SO(3)$ that the $SO(1,3)$ is broken into in FRW (just like its real counterpart, $A_0$).

\section{VDM Solitons}
\label{sec:VDMHalos}

Once formed, the self-gravitating VDM structures should settle into equilibrium configurations. These structures are comprised of localized cores whose size is determined by the de-Broglie/Jeans wavelength and an outer envelope, indistinguishable from CDM structures. The cores are stationary solutions of the Schr\"{o}dinger-Poisson system (commonly referred to as solitons). We now look for such solutions (after ignoring the expansion of the universe). 

We are primarily interested in the ground state soliton solutions of eqs. \eqref{eq:PoissonFinal} and \eqref{eq:EOMSchrjFinal}, which are (classically) stable and have a spherically symmetric distribution of energy. In other words, we look for $\Aj({\bf x},t)$ which give us a spherically symmetric, static gravitational, potential $\Phi=\Phi(|{\bf x}|)$.\footnote{Spherically symmetric mass distributions imply a vanishing angular momentum, which is expected for halos formed from cosmological initial conditions.} They represent the cores of VDM halos. However, we also discuss cylindrical and planar soliton solutions, which can comprise the cores of filamentary and planar VDM structures.

The $\Aj({\bf x},t)$ solitons spontaneously break the internal $SU(3)$ symmetry, eq.\ \eqref{eq:SU3Transf}. Since $\Ao({\bf x},t)$ is not $SU(3)$ invariant, but it transforms as a scalar under spatial coordinate rotations, we use it to classify our stationary ground state solutions with a spherically symmetric energy distribution. 

We divide the solutions into three categories. The first one corresponds to $\Ao({\bf x},t)=0$, which implies a divergence-free $\Aj$, according to the Lorenz constraint, eq. \eqref{eq:LorenzFinal}. The second category of soliton solutions corresponds to a spherically symmetric $\Ao=\Ao(|{\bf x}|,t)$. As we show below, these $\Aj$ solitons are reminiscent of the hedgehog vector solutions in the context of topological defects in 3-dimensions. The third category of solitons with a spherically symmetric energy distribution corresponds to a generic ${\bf x}$-dependence of $\Ao=\Ao({\bf x},t)$. 

In the following subsections we consider isolated configurations for which $\Aj({\bf x},t)$ and $\Phi(|{\bf x}|)$ are non-singular near the origin and approach zero at $|{\bf x}|\rightarrow \infty$. The net mass of each $\Aj$ is given by $M^{\left(j\right)}=m\int d^3x |\Aj|^2$ and the total mass of the soliton is $M_{\rm tot}=\sum_j M^{\left(j\right)}$. The gravitational potential energy of each $\Aj$ is $W^{\left(j\right)}=(1/2)\int d^3x \Phi m|\Aj|^2$ and the total gravitational potential energy is $W_{\rm tot}=\sum_j W^{\left(j\right)}$. These considerations do not apply to Sections \ref{sec:HedgeFilament}, \ref{sec:Filament} and \ref{sec:Walls} where we briefly study objects with cylindrical and planar symmetry.

We use the dimensionless quantities
\Beq
\label{eq:RescVar}
\tilde{\mathcal{A}}^j&=\frac{m_{\rm Pl}^3}{mM_{\rm tot}^2}\frac{\Aj}{m^{3/2}}\,,\qquad\tilde{\Phi}=\frac{m_{\rm Pl}^4}{m^2M_{\rm tot}^2}\Phi\,,\\\tilde{W}^{(j)}&=\frac{m_{\rm Pl}^4}{M_{\rm tot}^2 m^2}\frac{W^{(j)}}{M_{\rm tot}}\,,\,\quad\tilde{x}^{\mu}=\frac{m^2}{m_{Pl}^2}x^{\mu}M_{\rm tot}\,.
\Eeq
In terms of these new variables the fraction of energy stored in each $\Aj$ is given by the norm of the rescaled wavefunction $\tilde{M}^{\left(j\right)} \equiv \int d^3\tilde{x}|\tilde{\mathcal{A}}^j|^2=M^{\left(j\right)}/M_{\rm tot}\leq1$, with $\sum_j\tilde{M}^{(j)}=1$.

\begin{figure}[t] 
   \centering
   \includegraphics[width=3.45in]{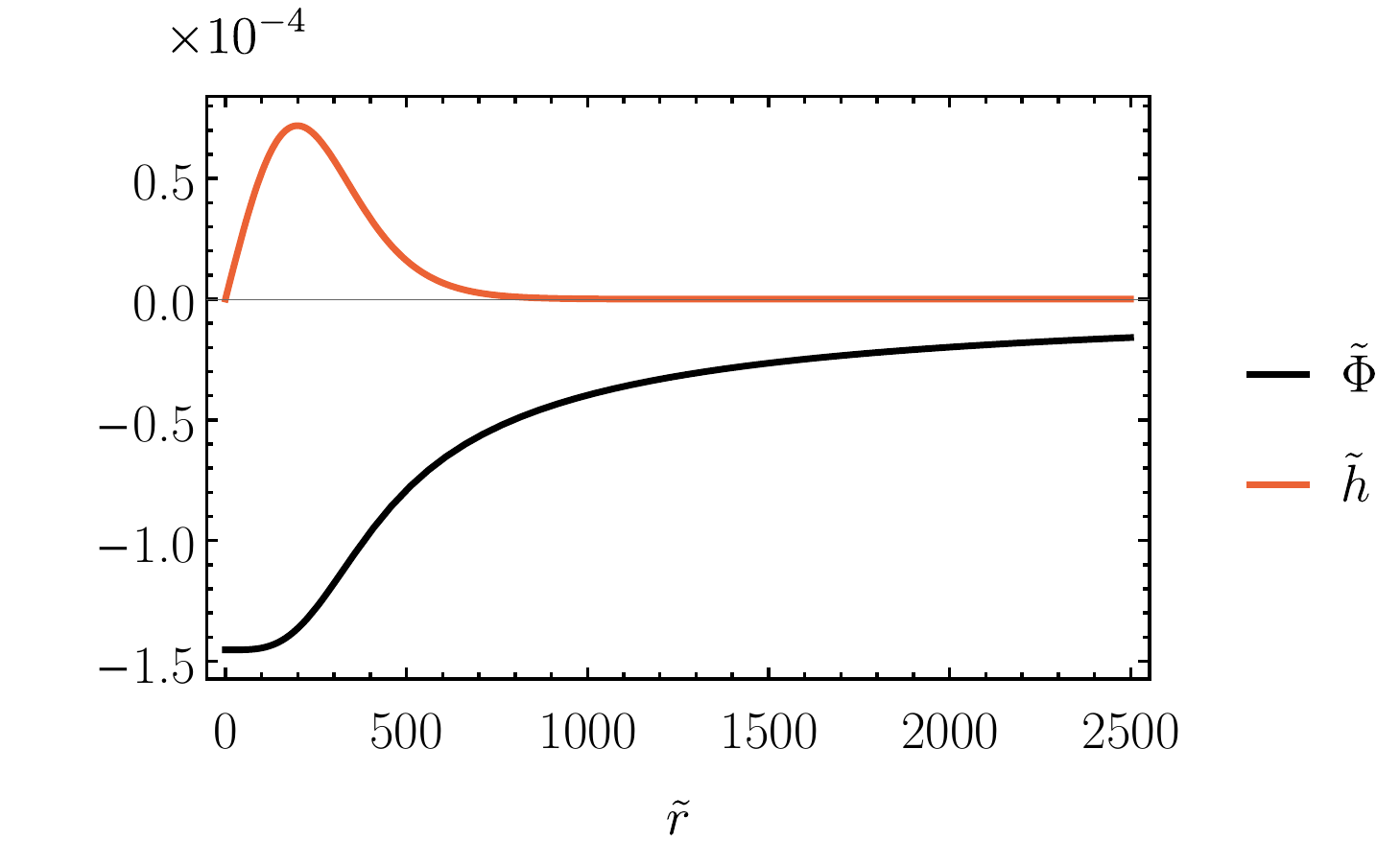}
   \caption{The spherical hedgehog solution, $\Aj=(\mathcal{A}^r,\mathcal{A}^{\theta},\mathcal{A}^{\varphi})=(h(|{\bf x}|)e^{-imE_ht},0,0)$, given in terms of the rescaled quantities from eq.\ \eqref{eq:RescVar}. The solution is normalized so that $\int \tilde{h}^2(\tilde{r})4\pi\tilde{r}^2dr=1$.}
   \label{fig:Hedgehog}
\end{figure}

\subsection{$\Ao=0$ solitons: non-spherical halos}

We begin with the most general ansatz for a transverse  $\Ai$
\Beq
\label{eq:A0zeroAnsatz}
\Ai({\bf x},t)=\epsilon^{ijk}\partial_jz^k({\bf x},t)\,,
\Eeq
where $z^j({\bf x},t)$ is a complex 3-component vector. Note that the Lorenz constraint, eq.\ \eqref{eq:LorenzFinal}, implies $im\Ao=\partial_i \Ai({\bf x},t)=0$..

The energy density, $T^0{}_0({\bf x},t)$, for this configuration is proportional to
\Beq
\label{eq:A0zeroDensity}
(\Aj{}^{*}\Aj)({\bf x},t)=\partial_i z^{k*} \partial_i z^k-\partial_i z^{k*} \partial_k z^i\,.
\Eeq
The only way to get a spherically-symmetric, static energy distribution, $T^0{}_0=T^0{}_0(|{\bf x}|)$ is to have a $z^j$ with $z^r\neq0$ and $z^{\theta}=z^{\varphi}=0$. However, this makes the right hand side in eq.\ \eqref{eq:A0zeroDensity}, vanish. 

Hence, there are no solitons with spherically symmetric and static energy distribution, $T^0{}_0(|{\bf x}|)$, and vanishing $\Ao$. The last condition may be met by solitons with non-spherically symmetric $T^0{}_{0}$.

\subsection{$\Ao=\Ao(|{\bf x}|,t)$ solitons: excited hedgehog halos}
\label{sec:Hedgehog}

We now consider the general trial solution 
\Beq
\label{eq:A0nonzeroAnsatz}
\Aj({\bf x},t)=z^j({\bf x},t)\,.
\Eeq
The Lorenz constraint for this configuration, eq.\ \eqref{eq:LorenzFinal}, reads
\Beq
\label{eq:A0LorenzA0nonzero}
\Ao=\Ao({\bf x},t)=-\frac{i}{m}\partial_j z^j({\bf x},t)\,
\Eeq
while the stress tensor is proportional to the inner product
\Beq
\label{eq:A0nonzeroDensity}
(\Aj{}^{*}\Aj)({\bf x},t)=(z^{j*}z^j)({\bf x},t)\,.
\Eeq
Eq.\ \eqref{eq:A0nonzeroDensity} implies that a spherically-symmetric energy distribution requires the complex 3-component vector from eq.\ \eqref{eq:A0nonzeroAnsatz} to have a spherically-symmetric coordinate dependence, i.e., $z^j=z^j(|{\bf x}|,t)$, up to arbitrary phases.

The phases drop out if we wish to have a spherically-symmetric $\Ao=\Ao(|{\bf x}|,t)$, see eq.\ \eqref{eq:A0LorenzA0nonzero}. This further leads to $z^r\neq0$, and $z^\theta = 0$, while the remaining components of the complex vector, $z^{\varphi}$ are unconstrained.\footnote{Recall that in spherical polars $\partial_j z^j({\bf x},t) = \frac{1}{r^2}\partial_r (r^2 z^r)+ \frac{1}{r \sin\theta}\partial_\theta(z^\theta \sin\theta) + \frac{1}{r \sin\theta}\partial_\varphi z^\varphi$.} We now consider the simplest scenario in which $z^{\varphi}=0$. This leaves us with a `hedgehog' solution, 
\Beq
\label{eq:Hedgehog}
\Aj=(\mathcal{A}^r,\mathcal{A}^{\theta},\mathcal{A}^{\varphi})=(h(|{\bf x}|)e^{-imE_ht},0,0)\,,
\Eeq
where we assume the stationary solution $z^r=h(|{\bf x}|)e^{-imE_ht}$. Note that $|E_h|\ll1$, since we are in the non-relativistic regime. This $h(|{\bf x}|)$ is governed by the Schr\"{o}dinger-Poisson equations 
\Beq
\label{eq:ShcrPoisHedg}
\left[-\frac{1}{2r^2}\frac{d}{dr}\left(r^2\frac{d}{dr}\right)+\frac{2}{r^2} \right]h+m\Phi h&=mE_h h\,,\\
\frac{1}{r^2}\frac{d}{dr}\left(r^2\frac{d\Phi}{dr}\right)&=\frac{m|h|^2}{2 m_{\rm Pl}^2}\,.
\Eeq
The last term in the square brackets is a consequence of the fact that we are considering the radial component of the vector dark matter field and it is not present in the case of scalar FDM. Furthermore, unlike the scalar FDM case, here we are forced to set the wavefunction to zero at the origin, $\mathcal{A}^r(|{\bf x}|=0,t)=h(|{\bf x}|=0,t)=0$, since this is the only way to make the vector field non-singular (i.e., continuous) there. Therefore, the hedgehog solution has a node at $r=0$ and is unlikely to be the ground-state VDM halo. In Fig.\ \ref{fig:Hedgehog}, we show the hedgehog solution with lowest energy (Appendix \ref{App:HHNum} contains the details of the numerical procedure used to solve eq.\ \eqref{eq:ShcrPoisHedg}). Higher-energy hedgehog solutions have more zero crossings of $h(r)$. The gravitational potential energy of the lowest-energy hedgehog solution is $\tilde{W}_{\rm hedg}=-5.71\times10^{-5}$ and is greater than the one of the axion FDM ground-state soliton, $\tilde{W}_{\rm axion}=-1.72\times10^{-4}$ \cite{Hui:2016ltb}. However, the first excited state of the axion FDM soliton has a gravitational binding $\tilde{W}_{\rm axion}^{n=1}=-3.25\times10^{-5}$ \cite{Hui:2016ltb}, which is weaker than the one of the hedgehog ground-state.

The simplest spherically-symmetric halo with non-zero $\Ao(|{\bf x}|,t)$ and only one non-vanishing spatial-vector component,  $\mathcal{A}^r$, therefore has a hedgehog-like solution, eq.\ \eqref{eq:Hedgehog} which is less gravitationally bound than the scalar FDM ground-state soliton solution. This is an indication that we have to go beyond these assumptions to find the ground-state VDM soliton. In the next section we show that configurations with non-spherically-symmetric $\Ao({\bf x},t)$ having all three $\Aj$ components excited can have a gravitational binding equal to the one of the FDM ground-state soliton, as long as they have the same temporal dependence. 

\begin{figure}[t] 
   \centering
   \includegraphics[width=3.45in]{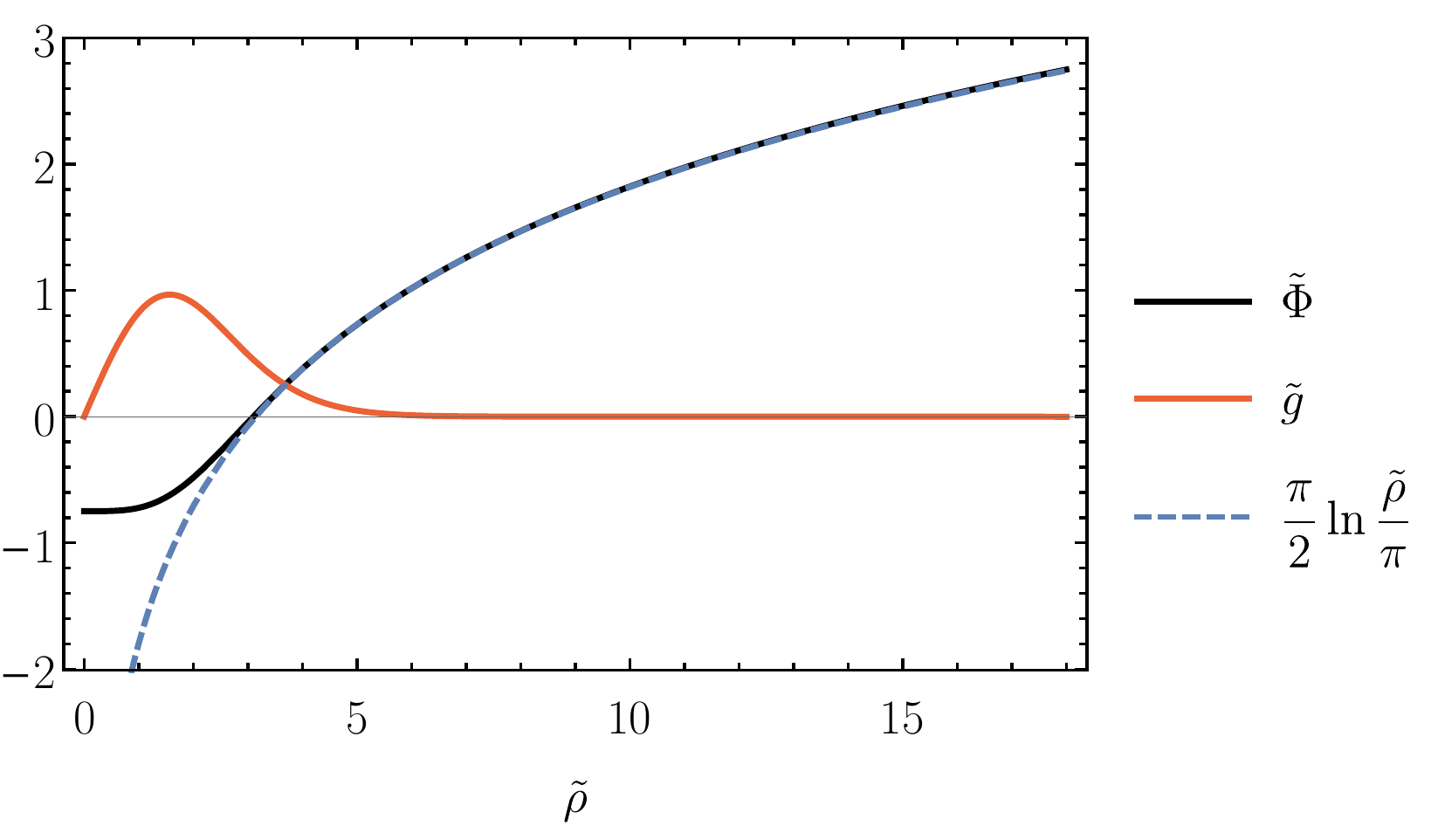}
   \caption{The hedgehog cylindrical solution, $\Aj=(\mathcal{A}^\rho,\mathcal{A}^{\theta},\mathcal{A}^{z})=(g(\rho)e^{-imE_gt},0,0)$, given in terms of the rescaled quantities from eq.\ \eqref{eq:RescVar}. The solution is normalized so that $\tilde{g}'(\tilde{\rho}=0)=1$.}
   \label{fig:HedgehogCyl}
\end{figure}

\subsubsection{Cylindrical hedgehog}
\label{sec:HedgeFilament}

Before we move onto study spherical solutions with more general spatial dependence, we remark upon another set of symmetric field solutions that are unique to vector fields: those with cylindrical symmetry
\begin{align}
\label{eq:HedgecylAnst}
\Aj=(\mathcal{A}^\rho,\mathcal{A}^{\theta},\mathcal{A}^{z})=(g(\rho)e^{-imE_gt},0,0),
\end{align}
where $\rho$ is the cylindrical radial coordinate. These solutions correspond to extended, string-like configurations and are solutions to the Schr\"{o}dinger-Poisson equations
\Beq
\label{eq:ShcrPoiscyl}
\left[-\frac{1}{2\rho}\frac{d}{d\rho}\left(\rho\frac{d}{d\rho}\right)+\frac{1}{\rho^2} \right]g+m\Phi g&=mE_g g\,,\\
\frac{1}{\rho}\frac{d}{d\rho}\left(\rho\frac{d\Phi}{d\rho}\right)&=\frac{m|g|^2}{2 m_{\rm Pl}^2}\,.
\Eeq
Solutions of these equations are again forced to vanish at $\rho = 0$, $g(\rho = 0) = 0$, and lead to gravitational potentials with $\ln(\rho)$ dependence at large $\rho$, see Fig.\ \ref{fig:HedgehogCyl}.  These are not ground state solitons, and we relegate further discussion of these solutions to Appendix \ref{App:ggnum}. See also Section \ref{sec:Filament} for the ground-state VDM filaments. Note that the last term in the square brackets in eq.\ \eqref{eq:ShcrPoiscyl} is not present in the cylindrical limit of scalar FDM and thus the solutions of eq.\ \eqref{eq:ShcrPoiscyl} are different from the scalar field case \cite{tan1990self,wang2006numerical}. However, they could be relevant to filamentary VDM structures observed on galactic and cosmic scales, with hedgehog boundary conditions.  We note that an identical energy distribution to the hedgehog cylinder can be obtained if we assume the azimuthal ansatz $\Aj=(\mathcal{A}^\rho,\mathcal{A}^{\theta},\mathcal{A}^{z})=(0,g(\rho)e^{-imE_gt},0)$, where $g(\rho)$ and $E_g$ are the ones from eq.\ \eqref{eq:ShcrPoiscyl}.

\subsection{$\Ao=\Ao({\bf x},t)$ solitons: ground state halos}

The ansatz in eq.\ \eqref{eq:A0nonzeroAnsatz} and the subsequent analysis in eqs. \eqref{eq:A0LorenzA0nonzero} and \eqref{eq:A0nonzeroDensity} also applies to spherically-symmetric mass distributions with $\Ao({\bf x},t)$ having a generic ${\bf x}$-dependence. In particular, the complex vector function in eq.\ \eqref{eq:A0nonzeroAnsatz} is still $z^j=z^j(|{\bf x}|,t)$, in order to have a spherically-symmetric energy density, according to eq.\ \eqref{eq:A0nonzeroDensity}. However, this time there are no constraints on the forms of the individual components of $z^j$, since $\Ao({\bf x},t)$ has a generic ${\bf x}$-dependence

\begin{figure}[t] 
   \centering
   \includegraphics[width=3.45in]{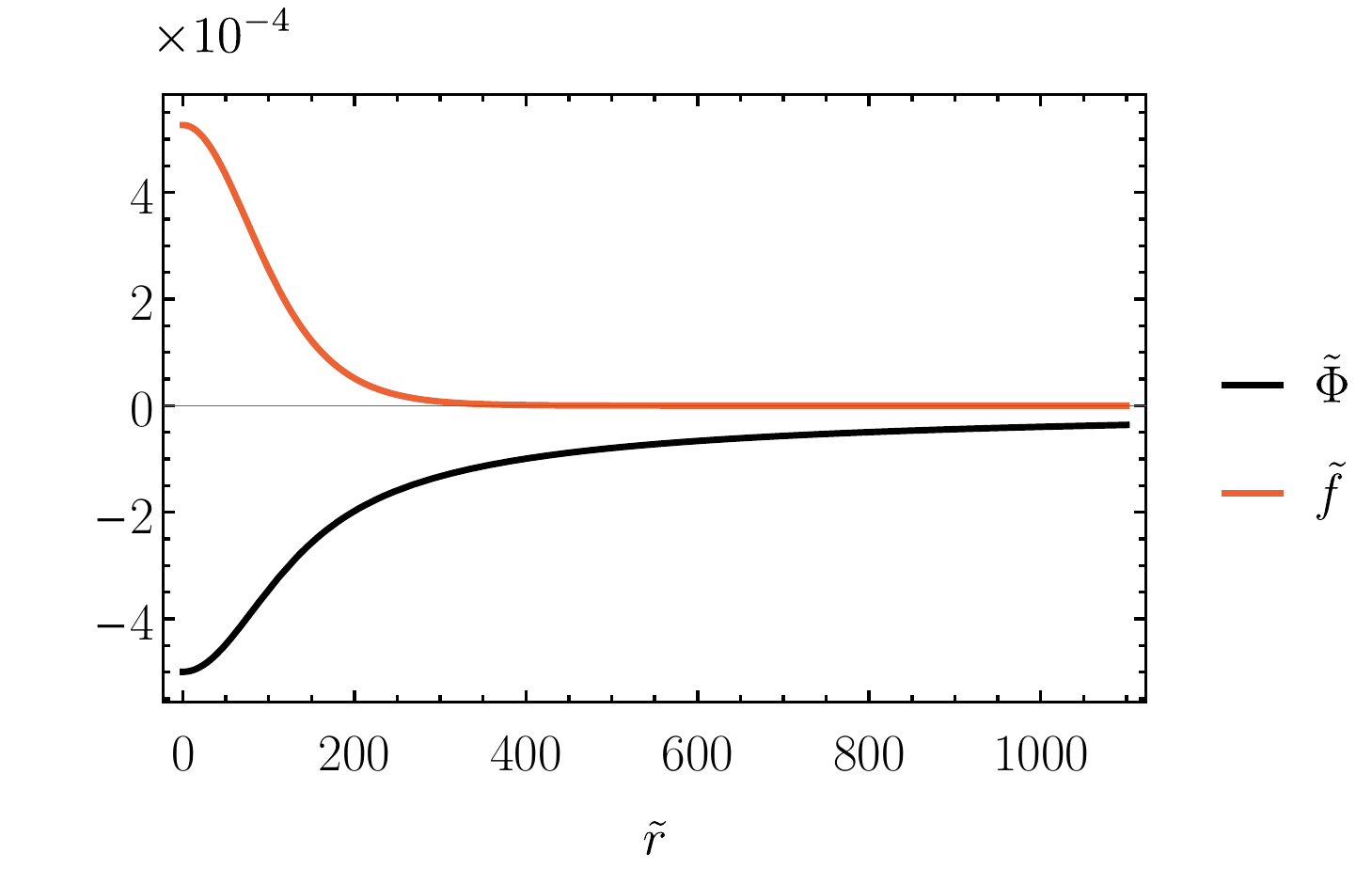}
   \caption{The ground-state Cartesian VDM soliton, $\Aj=(\mathcal{A}^x,\mathcal{A}^y,\mathcal{A}^z)=f(|{\bf x}|)e^{-imEt}(w^x,w^y,w^z)$, for arbitrary $w^j$, given in terms of the rescaled quantities from eq.\ \eqref{eq:RescVar}. The $\tilde{f}(\tilde{r})$ and $\tilde{\Phi}(\tilde{r})$ profiles are identical to the ones of the scalar FDM ground state soliton solution, see, e.g., Ref. \cite{Hui:2016ltb}. We use the same normalization as in Fig.\ \ref{fig:Hedgehog}.}
   \label{fig:Soliton}
\end{figure}

This simplest ansatz which respects these conditions is 
\Beq
\label{eq:zCart}
z^j(|{\bf x}|,t)=w^jf(|{\bf x}|)e^{-imEt}\,,
\Eeq
where $w^j$ is a constant vector with unit norm, $w^{j*}w^j=1$, (wlog). Hence, the Cartesian components of VDM can be written as
\Beq
\label{eq:Cartesian}
\Aj=(\mathcal{A}^x,\mathcal{A}^y,\mathcal{A}^z)=f(|{\bf x}|)e^{-imEt}(w^x,w^y,w^z)\,,
\Eeq
and are solutions of
\Beq
\label{eq:ShcrPoisCart}
-\frac{1}{2r^2}\frac{d}{dr}\left(r^2\frac{d}{dr}\right)f+m\Phi f&=mE f\,,\\
\frac{1}{r^2}\frac{d}{dr}\left(r^2\frac{d\Phi}{dr}\right)&=\frac{m|f|^2}{2 m_{\rm Pl}^2}\,.
\Eeq
$f(|{\bf x}|)$ is governed by the same Schr\"{o}dinger-Poisson system as the wavefunction of the axion FDM field \cite{Hui:2016ltb}. This time we can set $f(r=0)\neq0$, (akin to the axion FDM halo case and unlike the hedgehog case) since we are considering Cartesian vector components and there is no direction singularity at the origin. Therefore, the $f(|{\bf x}|)$ solutions are identical to the scalar FDM solitons. We focus on the ground-state soliton, see Fig.\ \ref{fig:Soliton}. We distinguish the different ground-state vector solutions, eq.\ \eqref{eq:Cartesian}, only by the choice of the constant complex vector $w^j$. Otherwise, they are degenerate. Their gravitational properties are identical for all $w^j$ and indistinguishable from the scalar FDM soliton, since they all have the same energy density distribution for the same total mass (thus $\tilde{W}_{\rm tot}=\tilde{W}^{(x)}+\tilde{W}^{(y)}+\tilde{W}^{(z)}=\tilde{W}_{\rm axion}=-1.72\times10^{-4}$). However, they are distinct from the hedgehog case in terms of mass distribution and are lower in gravitational energy, since the potential well is deeper at the origin for $f(r=0)\neq0$. We present the details of the numerical procedure to solve eq.\ \eqref{eq:ShcrPoisCart} in Appendix \ref{App:VDMGS}.

Hence, observational characteristics of the scalar and VDM ground-state solitons, such as the core gravitational potential, $\Phi(0)$, the half-mass radius,\footnote{Defined as $\sum_j\int_0^{\tilde{r}_{1/2}}d\tilde{r}4\pi\tilde{r}^2|\tilde{\Aj}(\tilde{r})|^2=1/2$.} $r_{1/2}$, and the virial velocity, $v_{\rm vir}\equiv\sqrt{-W_{\rm tot}/M_{\rm tot}}=(M_{\rm tot}m/m_{\rm Pl}^2)\sqrt{\tilde{W}_{\rm tot}}$, are identical for halos with the same total mass, $M_{\rm tot}$, and axions and vectors with the same particle mass, $m$. The properties of these soliton solutions are summarized in Table \ref{tab:tbl} along with the analogous properties of the lowest-energy hedgehog case.

Going beyond the ansatz in eq.\ \eqref{eq:zCart} is certainly possible (see, for example, Appendix \ref{App:GenericVDMGS}). However, it is unlikely that this will yield a VDM halo which is more strongly gravitationally bound than the axion FDM ground-state soliton. The breaking of the assumptions in eq.\ \eqref{eq:zCart} entails either having $\Aj$ components with (i) different $r$ dependences of the energy density profiles, $\propto|\Aj|^2$, and/or (ii) different time dependences, and/or (iii) different spatially-dependent phases (constant phases can be absorbed into $w^j$). None of these are expected to hold for the ground state. Case (i) can be refuted on symmetry grounds. Since the theory is $SU(3)$ invariant, its ground state can break the $SU(3)$ symmetry only spontaneously, i.e., there must be a set of degenerate ground states, related to each other via the $SU(3)$ transformation, eq.\ \eqref{eq:SU3Transf}. Hence, ground-state $\Aj$s must have the same $r$ dependence since their energy densities should $SU(3)$ rotate into each other. Unequal time-dependences are still allowed, i.e., $\Aj\propto e^{-imE^{(j)}t}$ with different $E^{(j)}$s. However, this implies a non-stationary $\Ao$ and thus case (ii) should not apply to the ground-state. Finally, the phases of the $\Aj$s should be spatially-independent (cf. case (iii)) for the ground-state VDM solitons. The velocity of the VDM superfluid (see Appendix \ref{App:SF}) is proportional to the spatial gradients of the phases. Thus a ground-state VDM halo with a vanishing fluid velocity must have $\Aj$s with spatially-constant phases.

\begin{table}[t]
\begin{tabular}{c|c|c|c}
$\Aj$& $\tilde{\Phi}(0)$ & $\tilde{r}_{1/2}$ & $\tilde{W}_{\rm tot}$ \\
\hline
{\rm Eq. \eqref{eq:Hedgehog}} & $-1.45\times10^{-4}$ & 304 & $-5.71\times10^{-5}$ \\
{\rm Eq. \eqref{eq:Cartesian}} & $-5.00\times10^{-4}$ & 98.6 & $-1.72\times10^{-4}$

\end{tabular}
\caption{Properties of the lowest-energy spherical hedgehog (first row) and Cartesian (second row) Vector Dark Matter soliton solutions. The dimensionless gravitational potential at the center of the halo, half-mass radius and gravitational binding energy are given in the second, third and fourth columns, respectively (see also eq.\ \eqref{eq:RescVar}). }
\label{tab:tbl}
\end{table}

\subsubsection{Filaments}
\label{sec:Filament}

\begin{figure}[t] 
   \centering
   \includegraphics[width=3.45in]{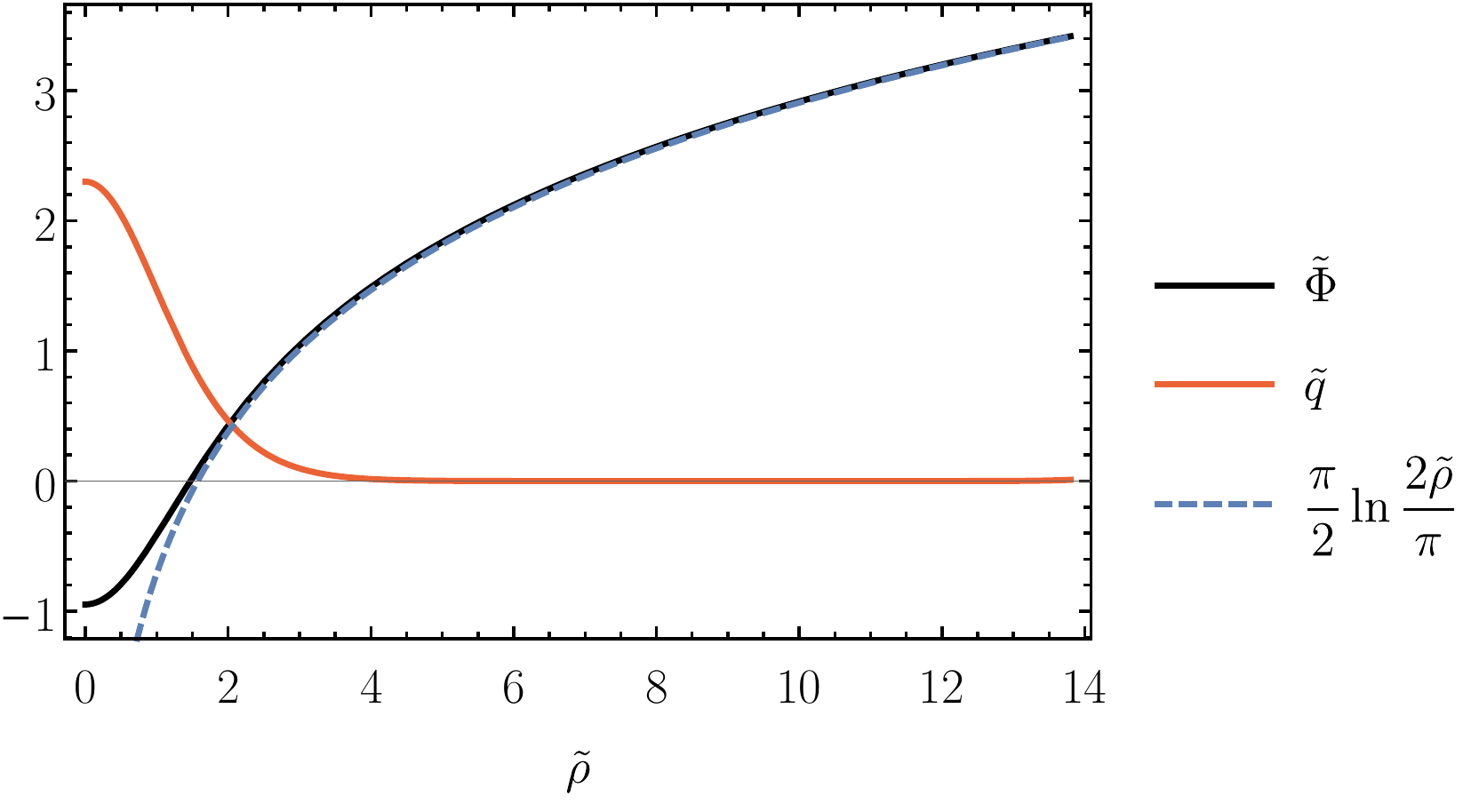}
   \caption{The lowest-energy cylindrical solution, $\Aj=(\mathcal{A}^x,\mathcal{A}^y,\mathcal{A}^{z})=q(\rho)e^{-imE_qt}(w^x,w^y,0)$, given in terms of the rescaled quantities from eq.\ \eqref{eq:RescVar}. The $\tilde{q}(\tilde{\rho})$ and $\tilde{\Phi}(\tilde{\rho})$ solutions are identical to the ones of the scalar FDM lowest-energy cylindrical solution, see, e.g., \cite{wang2006numerical}. The solution is normalized so that it gives rise to the same asymptotic behavior of the gravitational potential, $\tilde{\Phi}(\tilde{\rho}\rightarrow\infty)\rightarrow(\pi/2)\ln(\tilde{\rho})+{\rm const}$, as the one of the hedgehog string from Fig.\ \ref{fig:HedgehogCyl}.}
   \label{fig:Cyl}
\end{figure}

We  again briefly detour into solutions with cylindrically-symmetric distributions of mass. We consider the Cartesian components of the vector field
\begin{align}
\label{eq:CylAnsatz}
\Aj=(\mathcal{A}^x,\mathcal{A}^y,\mathcal{A}^{z})=q(\rho)e^{-imE_qt}(w^x,w^y,0)\,.
\end{align}
These solutions (like the ones from Section \ref{sec:HedgeFilament}) represent string-like configurations. They are solutions to the Schr\"{o}dinger-Poisson equations
\Beq
\label{eq:ShcrPoiscylCart}
-\frac{1}{2\rho}\frac{d}{d\rho}\left(\rho\frac{d}{d\rho}\right)q+m\Phi q&=mE_q q\,,\\
\frac{1}{\rho}\frac{d}{d\rho}\left(\rho\frac{d\Phi}{d\rho}\right)&=\frac{m|q|^2}{2 m_{\rm Pl}^2}\,,
\Eeq
where $\rho$ is the cylindrical radial coordinate. The same system of equations describes string-like configurations of scalar FDM \cite{tan1990self,wang2006numerical}. Furthermore, just like in the scalar field case, the solutions of these equations can have $q(\rho = 0) \neq 0$. Hence, the ground-state solution of the cylindrically symmetric Cartesian components of the vector field, eq.\ \eqref{eq:ShcrPoiscylCart}, is identical to the one of scalar FDM. It has a gravitational potential with $\ln(\rho)$ dependence at large $\rho$, see Fig.\ \ref{fig:Cyl} and Appendix \ref{App:cartggnum} where we explain how to obtain the solution numerically. The depth of the potential well at $\rho=0$ is greater than the one of the hedgehog filament from Section \ref{sec:HedgeFilament} (see end of Appendix \ref{App:cartggnum} for details), which suggests that the Cartesian case, eq.\ \eqref{eq:ShcrPoiscylCart}, gives rise to a stronger gravitational binding. Just like for spherical halos, when we have a cylindrical structure (filament), the lowest energy VDM solutions are the ones that are identical to the ground-state scalar FDM (cylindrical) soliton.

\subsubsection{Walls}
\label{sec:Walls}

For completeness we discuss solutions with mass distributions which have a planar symmetry. We consider a vector field with Cartesian components
\begin{align}
\Aj=(\mathcal{A}^x,\mathcal{A}^y,\mathcal{A}^{z})=p(x)e^{-imE_pt}(1,0,0)\,.
\end{align}
These solutions represent wall-like configurations lying in the $y$-$z$ plane. They are solutions to the Schr\"{o}dinger-Poisson equations
\Beq
\label{eq:ShcrPoisWall}
-\frac{1}{2}\frac{d^2p}{dx^2}+m\Phi p&=mE_p p\,,\\
\frac{d^2\Phi}{dx^2}&=\frac{m|p|^2}{2 m_{\rm Pl}^2}\,.
\Eeq
Wall-like configurations of scalar FDM are described by the same equations. Depending on the topology of the vector field we have two symmetry possibilities (i) when $p(x)=-p(-x)$, i.e., the vector field points in opposite directions on both sides of the plane, we are forced to have $p(x=0)=0$ and (ii) when $p(x)=p(-x)$, i.e., $\Aj$ always has the same direction, the vector field does not have to vanish in the plane of symmetry. Both cases correspond to one-dimensional solutions for scalar FDM \cite{1991JAP....70.2734S}, with (ii) corresponding to even solutions, including the nodeless ground-state, and (i) capturing odd excited solutions. 

The Large Scale Structure of the universe reveals that dark matter forms self-gravitating spherically-symmetric objects (i.e., halos), string-like structures (also known as filaments), and wall-like structures (often referred to as `pancakes'), see, for example, Ref.\ \cite{Bond:1995yt}. Unless there is a (topological) reason which forces the vector field to vanish in the center, on the axis or plane of symmetry, respectively, the ground-state VDM halos, filaments and walls are indistinguishable from the scalar FDM ones.

The reason why we have Cartesian vector solitons is because in the non-relativistic regime, $k\ll(am)$, the pressure of the vector field is negligible. In this limit the energy momentum tensor, eq. \eqref{eq:EMtensor}, loses its dependence on $F_{ij}$ and reduces to eq. \eqref{eq:EMNonR}. Anisotropic vector field configurations like the Cartesian solitons do not give rise to anisotropic pressure. This picture changes once we have relativistic field configurations, since the stresses induced by $F_{ij}$ are sizable. The only known way to realize spherically symmetric stable strongly gravitating solitons is to employ hedgehog solutions \cite{Brito:2015pxa}. The transition between these two regimes sets an upper bound on the Cartesian spherical solitons. In the non-relativistic limit $|\Phi(0)|\ll1$, implying $M_{\rm tot}\ll |\tilde{\Phi}(0)|^{-1/2}m_{\rm Pl}^2/m\sim10^{12}M_{\odot}(10^{-22}{\rm eV}/m)$, see Table \ref{tab:tbl} and eq. \eqref{eq:RescVar}. The bound is identical to the one for scalar FDM halos \cite{Hui:2016ltb}. The lower bound on the Cartesian VDM halos also coincides with the one for scalar FDM \cite{Hui:2016ltb}. Just like in the scalar case it can be derived in two independent ways, namely by either assuming that the mean density of the soliton, $\rho_{1/2}\sim M_{\rm tot}/(4\pi r_{1/2}^3/3)$, is greater than the virial density, $\sim200 \rho_{\rm crit}$, $\rho_{\rm crit}=3H^2m_{\rm Pl}^2$, or that $M_{\rm tot}$ has to be greater than the critical Jeans mass\footnote{Just like for scalar FDM \cite{Hu:2000ke}, the critical VDM Jeans scale, $\lambda_{\rm J}\sim v \sqrt{m_{\rm Pl}^2/\rho_{\rm crit}}$, is determined by equating it to the de Broglie scale, $\lambda_{\rm dB}\sim1/(mv)$, where $v$ is the non-relativistic velocity of a vector particle. Thus, we have $\lambda_{\rm J}\sim(m_{\rm Pl}^2/(m^2\rho_{\rm crit}))^{1/4}$.}, $M_{\rm J}\sim \rho_{\rm crit} \lambda_{\rm J}^3$. Both methods give roughly the same lower bound on the mass of the Cartesian spherical VDM soliton, $M_{\rm tot}>10^{7}M_{\odot}(10^{-22}{\rm eV}/m)^{3/2}$. Hence, $m\sim10^{-22}$ eV for VDM to explain the observed suppression of halos below $10^{7}-10^{8}M_{\odot}$, just like scalar FDM \cite{Hui:2016ltb}.

\section{Conclusions}
\label{sec:Conclusions}

In this work, we  have studied the non-relativistic limit of a massive vector field in a perturbed FRW spacetime. We explicitly showed that in this regime, the Cartesian dynamical degrees of freedom ($A^x$,$A^y$,$A^z$) behave as three copies of a massive scalar field that interact only gravitationally via the Poisson equation. We then focused on stationary, spatially-localized solutions of the non-relativistic theory---soliton solutions. The ground-state solitons we found have identical properties to the standard scalar FDM solitons, regardless of the distribution of energy among the $A^j$ fields. Scalar FDM halos and the resulting Fuzzy VDM halos are observationally indistinguishable on the basis of the gravitational properties of their cores in Newtonian limit.

We also found a novel class of self-gravitating solutions peculiar to VDM (i.e., not observed with scalar FDM), to which we refer to as hedgehog solitons,\footnote{Note that similar `Proca star' solutions have been studied in the context of a relativistic complex Proca field in Ref. \cite{Brito:2015pxa}.}  since the vector field topology is reminiscent of the one of the hedgehog solution describing the t'Hooft-Polyakov monopole \cite{tHooft:1974kcl,Polyakov:1974ek}. They involved the spherical or cylindrical components of the vector field, ($A^r$, $A^\theta$, $A^\varphi$) or ($A^r$, $A^\theta$, $A^z$), respectively. The lowest-energy hedgehog solutions lie between the ground and the first excited states of the Cartesian VDM solitons.

Lastly, we showed that there are no (exactly) spherical solutions of a massive vector field with a vanishing longitudinal component (but non-zero transverse components). 

The most salient distinction between axion and vector FDM remains the power spectrum of their initial fluctuations, which can affect the abundance of low mass halos \cite{Hu:2000ke,Lee:2020wfn} and depends on the primordial production mechanism. Misalignment straightforwardly produces the correct abundance of an axion with the theoretically desired mass and decay constant, $m\sim10^{-22}$ eV and $F\sim10^{17}$ GeV, respectively, in a miraculous fashion \cite{Hui:2016ltb}. However, when applied to a massive vector, production via misalignment requires non-minimal, highly-tuned couplings to gravity \cite{Arias:2012az}, which also lead to unitarity issues at low energies. Massive vectors minimally coupled to gravity can be produced by quantum fluctuations during inflation \cite{Graham:2015rva,Ema:2019yrd, Kolb:2020fwh, Ahmed:2020fhc}, but the catch is that the vector is too heavy, $m>10^{-5}$ eV, and can be a candidate for CDM, not FDM. Nevertheless, the vector power spectrum has a peak on intermediate scales, unlike the nearly scale-invariant scalar field spectrum from inflation. Production of ultra-light vectors, $m\gtrsim10^{-20}$ eV, can be realized with a resonant tachyonic decay of an oscillating axion \cite{Agrawal:2018vin, Dror:2018pdh, Bastero-Gil:2018uel, Co:2018lka} (which again gives rise to a strongly peaked VDM spectrum). We leave the study of the non-relativistic dynamics of Fuzzy VDM in such primordial embeddings for future work.

Another possible way to look for signatures of the vector nature of Dark Matter is to consider phenomena on Compton scales on which relativistic effects are not negligible. We plan to study such effects with simulations, carried out in the relativistic as well as non-relativistic regimes. Gravitational wave signatures arising from mergers and coalescing halos may have distinct signatures due to differences in the form of the stress tensor. Since the relativistic limits of VDM and scalar FDM are different, even perturbative post-Newtonian expansions should be sufficient to predict novel VDM signatures, which do not arise in scalar FDM models.

The additional freedom in the Schr\"{o}dinger-Poisson system we have derived here implies that there might be solutions beyond the simple-symmetric solutions we presented here. A non-relativistic simulation of the formation of a sufficiently large sample of VDM halos from cosmological initial conditions could reveal the whole spectrum of stable vector solitons. If novel vector solitons exist, their stability criteria may differ from the one for the scalar field solition \cite{Vakhitov1973}, which applies to the spherical vector soliton we found in this work.

Non-gravitational self-interactions of Fuzzy VDM could also potentially affect its soliton solutions (for a related study in the context of axion FDM, see Ref. \cite{Amin:2019ums}). Quartic vector self-interactions occur naturally in realizations of the Higgs and Stuckelberg mechanisms, albeit their relative strength needs to be suppressed for efficient primordial production of VDM \cite{Agrawal:2018vin}. Non-Abelian VDM also possesses non-gravitational self-interactions \cite{Nomura:2020zlm, Gross:2020zam}. We defer the investigation of such effects on the soliton formation and stability for future work.

\acknowledgments

We are grateful to Mustafa Amin,  Patrick Draper, and Zach Weiner for useful discussions. We additionally thank Mustafa Amin for suggesting we explore the soliton solutions. This work was supported in part by NASA Astrophysics Theory Grant NNX17AG48G.

\appendix

\section{Numerical VDM Soliton Solutions}
\label{App:Soliton}

In this Appendix, we provide details about the numerical procedure used to find the VDM halo solutions from Section \ref{sec:VDMHalos}.

We take the stationary-state ansatz, $\Aj({\bf x},t)=\psi^j({\bf x})e^{-imE^{\left(j\right)}t}$, which yields the time-independent Schr\"{o}dinger equations
\Beq
\label{eq:TimeIndSchr}
\left[-\frac{\nabla^2}{2m}+m\Phi({\bf x})\right]\psi^j({\bf x})=mE^{\left(j\right)}\psi^j({\bf x})\,.
\Eeq
Note that $|E^{\left(j\right)}|\ll1$ for the non-relativistic assumption to hold. We consider isolated configurations for which $\psi^j({\bf x})$ and $\Phi({\bf x})$ are regular near the origin and approach zero at $|{\bf x}|\rightarrow \infty$ (unless we drop the assumption of spherical symmetry and consider cylindrical symmetry instead). It follows directly from eq.\ \eqref{eq:TimeIndSchr} that $E^{\left(j\right)}$ is not the net energy per unit mass of the $j$th-component, since $M^{\left(j\right)}E^{\left(j\right)}=2W^{\left(j\right)}+\mathcal{K}^{\left(j\right)}+Q^{\left(j\right)}$, where $\mathcal{K}^{\left(j\right)}$ and $Q^{\left(j\right)}$ are interpreted as the kinetic and quantum energies of each $\Aj$ (see eq.\ \eqref{eq:VirialTheorem} for definitions), and the gravitational potential energy of each $\Aj$ is\footnote{$W^{\left(j\right)}\neq\mathcal{V}^{\left(j\right)}$ in general, cf. eq.\ \eqref{eq:VirialTheorem}.} $W^{\left(j\right)}=(1/2)\int d^3x \Phi m|\psi^j|^2$. For self-gravitating systems
\Beq
\Delta\Phi=\sum_j\frac{m|\psi^j|^2}{2m_{\rm Pl}^2}\,,
\Eeq
and $W_{\rm tot}=\sum_j W^{\left(j\right)}=\sum_j\mathcal{V}^{\left(j\right)}$ (see eq.\ \eqref{eq:VirialTheorem} for definitions). Moreover, the Virial Theorem from Appendix \ref{App:virial} implies for the stationary soliton $\sum_jM^{\left(j\right)}E^{\left(j\right)}=(3/2)W_{\rm tot}$.

In the rest of this appendix we will be working in terms of the rescaled dimensionless variables, following eq.\ \eqref{eq:RescVar},
\Beq
&\tilde{\psi}^j=\frac{m_{\rm Pl}^3}{mM_{\rm tot}^2}\frac{\psi^j}{m^{3/2}}\,,\qquad \tilde{\Phi}=\frac{m_{\rm Pl}^4}{m^2M_{\rm tot}^2}\Phi\,,\\
&\tilde{r}=\frac{m^2}{m_{Pl}^2}M_{\rm tot}|{\bf x}|\,,\qquad \tilde{E}^{\left(j\right)}=\frac{m_{\rm Pl}^4}{m^2M_{\rm tot}^2}E^{\left(j\right)}\,,
\Eeq
which reduce the time-independent Schr\"{o}dinger-Poisson system to
\Beq
\label{eq:EOMSchrjDimless}
-\frac{\tilde{\nabla}^2}{2}\tilde{\mathcal{\psi}}^j+\tilde{\Phi}\tilde{\mathcal{\psi}}^j=\tilde{E}^{(j)}\tilde{\psi^j}\,,\qquad\tilde{\nabla}^2\tilde{\Phi}=\frac{1}{2}\tilde{\psi}^{j*}\!\tilde{\psi}^j\,.
\Eeq

\subsection{VDM Hedgehog}
\label{App:HHNum}
We now find the numerical solutions for the hedgehog ansatz from eq.\ \eqref{eq:Hedgehog}, which we write as
\Beq
\label{eq:HedgehogDimLess}
\Aj&=(\mathcal{A}^r,\mathcal{A}^{\theta},\mathcal{A}^{\varphi})=(h(|{\bf x}|)e^{-imE_ht},0,0)\\
&=\frac{m^{1/2}m_{\rm Pl}}{M_{\rm tot}^2}(\tilde{h}(|\tilde{{\bf x}}|)e^{-i\tilde{E}_h\tilde{t}},0,0)\,.
\Eeq
The dimensionless time-independent Schr\"{o}dinger-Poisson equations, see eqs. \eqref{eq:ShcrPoisHedg} and \eqref{eq:EOMSchrjDimless}, are given by
\Beq
\label{eq:ShcrPoisHedgDimLess}
\left[-\frac{1}{2\tilde{r}^2}\frac{d}{d\tilde{r}}\left(\tilde{r}^2\frac{d}{d\tilde{r}}\right)+\frac{2}{\tilde{r}^2} \right]\tilde{h}+\tilde{\Phi} \tilde{h}&=\tilde{E}_h \tilde{h}\,,\\
\frac{1}{\tilde{r}^2}\frac{d}{d\tilde{r}}\left(\tilde{r}^2\frac{d\tilde{\Phi}}{d\tilde{r}}\right)&=\frac{\tilde{h}^2}{2}\,,
\Eeq
where we assume that $\tilde{h}(\tilde{r})$ has a spatially-independent complex phase and thus assumed wlog that it is real. 

The dimensionless Schr\"{o}dinger-Poisson system, eq.\ \eqref{eq:ShcrPoisHedgDimLess}, does not possess any known analytic solutions. To proceed, we solve the  system of equations in eq.\ \eqref{eq:ShcrPoisHedgDimLess} numerically using the `shooting' method. 

We first absorb the rescaled energy eigenvalue into the gravitational potential, $\tilde{\varphi}_h\equiv\tilde{\Phi}-\tilde{E}_h$. Then eq.\ \eqref{eq:ShcrPoisHedgDimLess} reduces to
\Beq
\label{eq:SPNumHedge}
\left[\frac{1}{2\tilde{r}^2}\frac{d}{d\tilde{r}}\left(\tilde{r}^2\frac{d}{d\tilde{r}}\right)-\frac{2}{\tilde{r}^2} \right]\tilde{h}&=\tilde{\varphi}_h\tilde{h}\,,\\
\frac{1}{\tilde{r}^2}\frac{d}{d\tilde{r}}\left(\tilde{r}^2\frac{d}{d\tilde{r}}\right)\tilde{\varphi}_h&=\frac{1}{2}h^2\,,
\Eeq
The system of differential equations, eq.\ \eqref{eq:SPNumHedge}, remains invariant under the rescaling
\Beq
\label{eq:RescHedge}
\tilde{r}\rightarrow\lambda\tilde{r}\,,\quad\tilde{h}(\tilde{r})\rightarrow\lambda^{-2}\tilde{h}(\lambda\tilde{r})\,,\quad\tilde{\varphi}_h(\tilde{r})\rightarrow\lambda^{-2}\tilde{\varphi}_h(\lambda\tilde{r})\,.
\Eeq
Under this rescaling the norm of the solutions transforms as $\int d^3\tilde{x} \tilde{h}^2(\tilde{r})\rightarrow \lambda^{-1}\int d^3\tilde{x} \tilde{h}^2(\tilde{r})$.

Next we integrate eq.\ \eqref{eq:SPNumHedge} and find a smooth and nodeless $\tilde{h}(\tilde{r})$ solution, linear near the origin, $\tilde{h}(\tilde{r})\propto \tilde{r}$, and approaching zero at $\tilde{r}\rightarrow\infty$, satisfying the normalization condition $\int d^3\tilde{x} \tilde{h}^2(\tilde{r})=1$. To do this we implement the following four-step algorithm:
\begin{enumerate}
\item We choose an arbitrary negative numerical value for $\tilde{\varphi}_h(0)$, put $\tilde{h}{}(0)=0$ and set $\tilde{\varphi}_h'(0)=0$ and $\tilde{h}{}'(0)=1$.
\item We integrate eq.\ \eqref{eq:SPNumHedge} from $\tilde{r}=0$ to $\infty$.
\item We keep changing $\tilde{\varphi}_h(0)$ and repeating Step 2 until we find a smooth and nodeless $\tilde{h}(\tilde{r})$ solution.
\item The smooth and nodeless solution obtained in Step 3 has $\int d^3\tilde{x} \tilde{h}^2(\tilde{r})=\lambda$. To normalize it, we rescale it according to eq.\ \eqref{eq:RescHedge}.
\end{enumerate}

The final solution from Step 4 is the lowest-energy hedgehog soliton, see Fig.\ \ref{fig:Hedgehog}. Once we have it, we can recover $\tilde{E}_h$ in two independent ways. The first one makes use of the Virial Theorem which implies that $\tilde{E}_h=-3\int d^3\tilde{x} \tilde{\varphi}_h(\tilde{{\bf x}})\tilde{h}^2(\tilde{{\bf x}})$. The second method is based on the assumption that $\tilde{\Phi}(\tilde{{\bf x}})$ vanishes at infinity. Then from the definition of $\tilde{\varphi}_h$ it follows that $\tilde{E}_h=-\tilde{\varphi}_h(\infty)$. Both methods yield results in agreement to arbitrary high-precision.\footnote{For the second method we also fitted $a\tilde{r}^{-1}+b$ to $\tilde{\varphi}_h(\tilde{r})$ at large distances, since $\tilde{\Phi}(\tilde{r})$ is expected to fall off as $\tilde{r}^{-1}$ far away from the spherically symmetric soliton.} 

\subsection{VDM hedgehog strings}
\label{App:ggnum}

Here we find the numerical solutions for the cylindrical-hedgehog ansatz from eq.\ \eqref{eq:HedgecylAnst}, which we write as
\Beq
\label{eq:HedgehogCylDimLess}
\Aj&=(\mathcal{A}^\rho,\mathcal{A}^{\theta},\mathcal{A}^{z})=(g(\rho)e^{-imE_gt},0,0)\\
&=\frac{m^{1/2}m_{\rm Pl}}{M_{\rm tot}^2}(\tilde{g}(\tilde{\rho})e^{-i\tilde{E}_g\tilde{t}},0,0)\,.
\Eeq
The dimensionless time-independent Schr\"{o}dinger-Poisson equations, see eqs. \eqref{eq:ShcrPoiscyl} and \eqref{eq:EOMSchrjDimless}, are
\Beq
\label{eq:ShcrPoisHedgCylDimLess}
\left[-\frac{1}{2\tilde{\rho}}\frac{d}{d\tilde{\rho}}\left(\tilde{\rho}\frac{d}{d\tilde{\rho}}\right)+\frac{1}{\tilde{\rho}^2} \right]\tilde{g}+\tilde{\Phi} \tilde{g}&=\tilde{E}_g \tilde{g}\,,\\
\frac{1}{\tilde{\rho}}\frac{d}{d\tilde{\rho}}\left(\tilde{\rho}\frac{d\tilde{\Phi}}{d\tilde{\rho}}\right)&=\frac{\tilde{g}^2}{2}\,,
\Eeq
where $\tilde{g}(\tilde{\rho})$ is assumed to have a spatially-independent complex phase which we can ignore. 

The dimensionless cylindrical Schr\"{o}dinger-Poisson system, eq.\ \eqref{eq:ShcrPoisHedgCylDimLess}, does not have analytic solutions. Thus, we again solve the  system of equations in eq.\ \eqref{eq:ShcrPoisHedgCylDimLess} numerically using the `shooting' method. 

We first absorb the rescaled energy eigenvalue into the gravitational potential, $\tilde{\varphi}_g\equiv\tilde{\Phi}-\tilde{E}_g$. Then eq.\ \eqref{eq:ShcrPoisHedgCylDimLess} reduces to
\Beq
\label{eq:SPNumCylHedge}
\left[\frac{1}{2\tilde{\rho}}\frac{d}{d\tilde{\rho}}\left(\tilde{\rho}\frac{d}{d\tilde{\rho}}\right)-\frac{1}{\tilde{\rho}^2} \right]\tilde{g}&=\tilde{\varphi}_g\tilde{g}\,,\\
\frac{1}{\tilde{\rho}}\frac{d}{d\tilde{\rho}}\left(\tilde{\rho}\frac{d}{d\tilde{\rho}}\right)\tilde{\varphi}_g&=\frac{1}{2}g^2\,.
\Eeq
The system of differential equations, eq.\ \eqref{eq:SPNumCylHedge}, is invariant under the rescaling
\Beq
\label{eq:RescCylHedge}
\tilde{\rho}\rightarrow\lambda\tilde{\rho}\,,\quad\tilde{g}(\tilde{\rho})\rightarrow\lambda^{-2}\tilde{g}(\lambda\tilde{\rho})\,,\quad\tilde{\varphi}_g(\tilde{\rho})\rightarrow\lambda^{-2}\tilde{\varphi}_g(\lambda\tilde{\rho})\,.
\Eeq

We now integrate eq.\ \eqref{eq:SPNumCylHedge} and find a smooth and nodeless $\tilde{g}(\tilde{\rho})$, linear near the origin, $g(\tilde{\rho})\propto \tilde{\rho}$, and approaching zero at $\tilde{\rho}\rightarrow\infty$. We do not consider the normalization of $\tilde{g}$ since it has an infinite extend along the $z$-direction. Instead, as a `normalization' condition we simply fix $\tilde{g}'(\tilde{\rho}=0)=1$. We next implement the following three-step algorithm:
\begin{enumerate}
\item We choose an arbitrary negative numerical value for $\tilde{\varphi}_g(0)$, put $\tilde{g}{}(0)=0$ and set $\tilde{\varphi}_g'(0)=0$ and $\tilde{g}{}'(0)=1$.
\item We integrate eq.\ \eqref{eq:SPNumCylHedge} from $\tilde{\rho}=0$ to $\infty$.
\item We keep changing $\tilde{\varphi}_g(0)$ and repeating Step 2 until we find a smooth and nodeless $\tilde{g}(\tilde{\rho})$ solution.
\end{enumerate}

The final solution from Step 3 is the lowest-energy hedgehog soliton, see Fig.\ \ref{fig:HedgehogCyl}. Once we have it, we do not recover $\tilde{E}_g$ since it can be absorbed into the constant appearing in the gravitational potential at infinity. Hence, we end up with $\tilde{\Phi}(\tilde{\rho})=\tilde{\varphi}_g(\tilde{\rho})$ which in the limit of $\tilde{\rho}\rightarrow\infty$ tends to $(\pi/2)\ln(\tilde{\rho}/\pi)$.

\subsection{VDM ground-state halos}
\label{App:VDMGS}

We now derive the numerical solutions of the ground-state ansatz from eq.\ \eqref{eq:Cartesian}, which we write as
\Beq
\label{eq:CartesianDimLess}
\Aj&=(\mathcal{A}^x,\mathcal{A}^y,\mathcal{A}^z)=f(|{\bf x}|)e^{-imEt}(w^x,w^y,w^z)\\
&=\frac{m^{1/2}m_{\rm Pl}}{M_{\rm tot}^2}\tilde{f}(|\tilde{\bf x}|)e^{-i\tilde{E}\tilde{t}}(w^x,w^y,w^z)\,.
\Eeq
The dimensionless time-independent Schr\"{o}dinger-Poisson equations, see eqs. \eqref{eq:ShcrPoisCart} and \eqref{eq:EOMSchrjDimless}, then become
\Beq
\label{eq:ShcrPoisCartDimLess}
-\frac{1}{2\tilde{r}^2}\frac{d}{d\tilde{r}}\left(\tilde{r}^2\frac{d}{d\tilde{r}}\right)\tilde{f}+\tilde{\Phi} \tilde{f}&=\tilde{E} \tilde{f}\,,\\
\frac{1}{\tilde{r}^2}\frac{d}{d\tilde{r}}\left(\tilde{r}^2\frac{d\tilde{\Phi}}{d\tilde{r}}\right)&=\frac{\tilde{f}^2}{2}\,,
\Eeq
where it is assumed that $\tilde{f}(\tilde{r})$ has a spatially-independent complex phase and hence, wlog, it is put to be real. 

The dimensionless Schr\"{o}dinger-Poisson system, eq.\ \eqref{eq:ShcrPoisCartDimLess}, is identical to the one for scalar-field FDM \cite{Hui:2016ltb}. It does not have any known analytic solutions, whereas its numerical solutions are well-known, see, e.g., Ref. \cite{Hui:2016ltb}. For completeness, we show how to obtain the known solutions of the system of equations, eq.\ \eqref{eq:ShcrPoisCartDimLess}, numerically using the `shooting' method. 

We repeat the procedure used for the hedgehog case. We begin by absorbing the rescaled energy eigenvalue into the dimensionless gravitational potential, $\tilde{\varphi}\equiv\tilde{\Phi}-\tilde{E}$. Then eq.\ \eqref{eq:ShcrPoisCartDimLess} becomes
\Beq
\label{eq:SPNumCart}
\frac{1}{2\tilde{r}^2}\frac{d}{d\tilde{r}}\left(\tilde{r}^2\frac{d}{d\tilde{r}}\right)\tilde{f}&=\tilde{\varphi}\tilde{f}\,,\\
\frac{1}{\tilde{r}^2}\frac{d}{d\tilde{r}}\left(\tilde{r}^2\frac{d}{d\tilde{r}}\right)\tilde{\varphi}&=\frac{1}{2}f^2\,.
\Eeq
Like in the hedgehog case, eq.\ \eqref{eq:SPNumCart} is invariant under the rescaling
\Beq
\label{eq:RescCart}
\tilde{r}\rightarrow\lambda\tilde{r}\,,\quad\tilde{f}(\tilde{r})\rightarrow\lambda^{-2}\tilde{f}(\lambda\tilde{r})\,,\quad\tilde{\varphi}(\tilde{r})\rightarrow\lambda^{-2}\tilde{\varphi}(\lambda\tilde{r})\,,
\Eeq
whereas the norm of the solutions transforms as $\int d^3\tilde{x} \tilde{f}^2(\tilde{r})\rightarrow \lambda^{-1}\int d^3\tilde{x} \tilde{f}^2(\tilde{r})$.

We proceed by numerically integrating eq.\ \eqref{eq:SPNumCart} and finding a smooth and nodeless $\tilde{f}(\tilde{r})$ solution, regular near the origin, and approaching zero at $\tilde{r}\rightarrow\infty$, normalized as follows $\int d^3\tilde{x} \tilde{f}^2(\tilde{r})=1$. We again implement a four-step algorithm:
\begin{enumerate}
\item We choose an arbitrary negative numerical value for $\tilde{\varphi}(0)$ and a positive one for $\tilde{f}{}(0)=0$, and set $\tilde{\varphi}'(0)$ and $\tilde{f}{}'(0)$ to zero.
\item We integrate eq.\ \eqref{eq:SPNumCart} from $\tilde{r}=0$ to $\infty$.
\item We keep changing $\tilde{\varphi}(0)$ and repeating Step 2 until we find a smooth and nodeless $\tilde{f}(\tilde{r})$ solution.
\item The smooth and nodeless solution obtained in Step 3 has $\int d^3\tilde{x} \tilde{f}^2(\tilde{r})=\lambda$. To normalize it, we rescale it according to eq.\ \eqref{eq:RescCart}.
\end{enumerate}

The final solution from Step 4 is the ground-state VDM soliton, see Fig.\ \ref{fig:Soliton}. We can again recover $\tilde{E}$ straightforwardly, in two independent ways, either by using the Virial Theorem identity $\tilde{E}=-3\int d^3\tilde{x} \tilde{\varphi}(\tilde{{\bf x}})\tilde{f}^2(\tilde{{\bf x}})$ or the assumption that $\tilde{\Phi}(\tilde{{\bf x}})$ vanishes at infinity and thus $\tilde{E}=-\tilde{\varphi}(\infty)$. The two approaches agree to arbitrary high-precision.\footnote{Like in the hedgehog case, for the second method we also fitted $a\tilde{r}^{-1}+b$ to $\tilde{\varphi}(\tilde{r})$ at large $\tilde{r}$, since we expect $\tilde{\Phi}(\tilde{r})\propto\tilde{r}^{-1}$ far away from the spherically symmetric soliton.} 

\subsection{VDM ground-state strings}
\label{App:cartggnum}

We now find the numerical solutions for the Cartesian cylindrical ansatz from eq.\ \eqref{eq:CylAnsatz}, which we write as
\Beq
\label{eq:CylDimLess}
\Aj&=(\mathcal{A}^x,\mathcal{A}^y,\mathcal{A}^{z})=q(\tilde{\rho})e^{-imE_qt}(w^x,w^y,0)\\
&=\frac{m^{1/2}m_{\rm Pl}}{M_{\rm tot}^2}\tilde{q}(\tilde{\rho})e^{-i\tilde{E}_q\tilde{t}}(w^x,w^y,0)\,.
\Eeq
The dimensionless time-independent Schr\"{o}dinger-Poisson equations, see eqs. \eqref{eq:ShcrPoiscylCart} and \eqref{eq:EOMSchrjDimless}, are
\Beq
\label{eq:ShcrPoisCylDimLess}
-\frac{1}{2\tilde{\rho}}\frac{d}{d\tilde{\rho}}\left(\tilde{\rho}\frac{d}{d\tilde{\rho}}\right)\tilde{q}+\tilde{\Phi} \tilde{q}&=\tilde{E}_q \tilde{q}\,,\\
\frac{1}{\tilde{\rho}}\frac{d}{d\tilde{\rho}}\left(\tilde{\rho}\frac{d\tilde{\Phi}}{d\tilde{\rho}}\right)&=\frac{\tilde{q}^2}{2}\,,
\Eeq
where $\tilde{q}(\tilde{\rho})$ is taken to be real wlog for the ground-state solution.

The dimensionless cylindrical Schr\"{o}dinger-Poisson system for the Cartesian components, eq.\ \eqref{eq:CylDimLess}, does not have known analytic solutions. We solve eq.\ \eqref{eq:ShcrPoisCylDimLess} numerically with the `shooting' method. 

As always, we absorb the rescaled energy eigenvalue into the gravitational potential, $\tilde{\varphi}_q\equiv\tilde{\Phi}-\tilde{E}_q$. Then eq.\ \eqref{eq:ShcrPoisCylDimLess} reduces to
\Beq
\label{eq:SPNumCyl}
\frac{1}{2\tilde{\rho}}\frac{d}{d\tilde{\rho}}\left(\tilde{\rho}\frac{d}{d\tilde{\rho}}\right)\tilde{q}&=\tilde{\varphi}_q\tilde{q}\,,\\
\frac{1}{\tilde{\rho}}\frac{d}{d\tilde{\rho}}\left(\tilde{\rho}\frac{d}{d\tilde{\rho}}\right)\tilde{\varphi}_q&=\frac{1}{2}q^2\,.
\Eeq
As before, the system of differential equations, eq.\ \eqref{eq:SPNumCyl}, is invariant under the rescaling
\Beq
\label{eq:RescCyl}
\tilde{\rho}\rightarrow\lambda\tilde{\rho}\,,\quad\tilde{q}(\tilde{\rho})\rightarrow\lambda^{-2}\tilde{q}(\lambda\tilde{\rho})\,,\quad\tilde{\varphi}_q(\tilde{\rho})\rightarrow\lambda^{-2}\tilde{\varphi}_q(\lambda\tilde{\rho})\,.
\Eeq

We now integrate eq.\ \eqref{eq:SPNumCyl} and find the smooth and nodeless ground-state $\tilde{q}(\tilde{\rho})$, regular near the origin and approaching zero at $\tilde{\rho}\rightarrow\infty$. We normalize $\tilde{q}$ in such a way that $\varphi_q(\tilde{\rho})= (\pi/2)\ln(\tilde{\rho})+{\rm const}$ at infinity. This allows us to compare the gravitational potential with the one of the hedgehog string from Appendix \ref{App:ggnum}, since the potential there has the same $\tilde{\rho}-$dependence at infinity, $(\pi/2)\ln\tilde{\rho}+{\rm const}$, and thus the filament has the same mass. We next implement the following four-step shooting algorithm:
\begin{enumerate}
\item We choose an arbitrary negative numerical value for $\tilde{\varphi}_q(0)$ and a positive one for $\tilde{q}{}(0)=0$, and set $\tilde{\varphi}_q'(0)$ and $\tilde{q}{}'(0)$ to zero.
\item We integrate eq.\ \eqref{eq:SPNumCyl} from $\tilde{\rho}=0$ to $\infty$.
\item We keep changing $\tilde{\varphi}_q(0)$ and repeating Step 2 until we find a smooth and nodeless $\tilde{q}(\tilde{\rho})$ solution.
\item We rescale the smooth and nodeless solution obtained in Step 3 according to eq.\ \eqref{eq:RescCyl} which gives $\varphi_q(\tilde{\rho})= (\pi/2)\ln(\tilde{\rho})+{\rm const}$ at infinity. The desired value of $\lambda$ needed for eq.\ \eqref{eq:RescCyl} is found by fitting.
\end{enumerate}

The final solution from Step 4 is the lowest-energy cylindrical soliton, see Fig.\ \ref{fig:Cyl}. We do not need to recover $\tilde{E}_q$ since in the cylindrical case it can be absorbed into the constant appearing in the gravitational potential at $\tilde{\rho}\rightarrow\infty$. Thus, we have $\tilde{\Phi}(\tilde{\rho})=\tilde{\varphi}_q(\tilde{\rho})$ which at $\tilde{\rho}\rightarrow\infty$ goes as $(\pi/2)\ln(2\tilde{\rho}/\pi)$. We then find from the numerical solutions that the difference in the depths of the gravitational potential wells (at $\tilde{\rho}=0$) of the hedgehog string from Appendix \ref{App:ggnum} and of the filament here is $\tilde{\varphi}_g(0)-\tilde{\varphi}_q(0)+(\pi/2)\ln(2)=(\pi/3)+3/(4\pi)$. Hence, the Cartesian filament, eq.\ \eqref{eq:CylDimLess}, has a deeper potential well than the hedgehog string, eq.\ \eqref{eq:HedgehogCylDimLess}.

\subsection{Generic VDM ground-state solutions}
\label{App:GenericVDMGS}

We show here that the ground-state VDM halos can be also obtained from a more generic ansatz than the one in eq.\ \eqref{eq:CartesianDimLess}, namely
\Beq
\label{eq:CartesianDimLess1}
\Aj&=\psi^j(|{\bf x}|)e^{-imE^{(j)}t}=(\mathcal{A}^x,\mathcal{A}^y,\mathcal{A}^z)\\
&=\frac{m^{1/2}m_{\rm Pl}}{M_{\rm tot}^2}\left(\tilde{\psi}^xe^{-i\tilde{E}^{(x)}\tilde{t}},\tilde{\psi}^ye^{-i\tilde{E}^{(y)}\tilde{t}},\tilde{\psi}^ze^{-i\tilde{E}^{(z)}\tilde{t}}\right)\,.
\Eeq
The dimensionless time-independent Schr\"{o}dinger-Poisson system, see eq.\ \eqref{eq:EOMSchrjDimless}, is
\Beq
\label{eq:ShcrPoisCartDimLessj}
-\frac{1}{2\tilde{r}^2}\frac{d}{d\tilde{r}}\left(\tilde{r}^2\frac{d}{d\tilde{r}}\right)\tilde{\psi}^j+\tilde{\Phi} \tilde{\psi}^j&=\tilde{E}^{(j)} \tilde{\psi}^j\,,\\
\frac{1}{\tilde{r}^2}\frac{d}{d\tilde{r}}\left(\tilde{r}^2\frac{d\tilde{\Phi}}{d\tilde{r}}\right)&=\frac{\tilde{\psi}^j\tilde{\psi}^j}{2}\,,
\Eeq
where it is assumed that the $\tilde{\psi}^{j}$s are real. 

This extended Schr\"{o}dinger-Poisson system does not possess any known analytic solutions (for any number of $j$-components). To make further progress, we make the simplifying assumption of $\tilde{E}^{(x)}=\tilde{E}^{(y)}=\tilde{E}^{(z)}=\tilde{E}$ and solve the  system of equations from eq.\ \eqref{eq:ShcrPoisCartDimLessj} with the aid of the `shooting' method.

We start by defining, $\tilde{\varphi}\equiv\tilde{\Phi}-\tilde{E}$. Then eq.\ \eqref{eq:ShcrPoisCartDimLessj} simplifies to
\Beq
\label{eq:SPNum}
\frac{1}{2\tilde{r}^2}\frac{d}{d\tilde{r}}\tilde{r}^2\frac{d\tilde{\psi}^j}{d\tilde{r}}=\tilde{\varphi}\tilde{\psi}^j\,,
\qquad\frac{1}{\tilde{r}^2}\frac{d}{d\tilde{r}}\tilde{r}^2\frac{d\tilde{\varphi}}{d\tilde{r}}=\frac{1}{2}\tilde{\psi}^{j}\tilde{\psi}^{j}\,.
\Eeq
Eq.\ \eqref{eq:SPNum} is invariant under
\Beq
\label{eq:Resc}
\tilde{r}\rightarrow\lambda\tilde{r}\,,\quad\tilde{\psi^j}(\tilde{r})\rightarrow\lambda^{-2}\tilde{\psi^j}(\lambda\tilde{r})\,,\quad\tilde{\varphi}(\tilde{r})\rightarrow\lambda^{-2}\tilde{\varphi}(\lambda\tilde{r})\,,
\Eeq
and the sum of the norms of the solutions transforms as $\sum_j\tilde{M}^{\left(j\right)}\rightarrow \lambda^{-1}\sum_j\tilde{M}^{\left(j\right)}$.

We continue by integrating eq.\ \eqref{eq:SPNum} and finding smooth and nodeless $\tilde{\psi}^j(\tilde{r})$ solutions, regular near the origin and approaching zero at $\tilde{r}\rightarrow\infty$, satisfying the normalization condition $\sum_j\tilde{M}^{\left(j\right)}=1$. To this end we use the following four-step algorithm:
\begin{enumerate}
\item We choose numerical values for $\tilde{\psi}^j(0)$ and $\tilde{\varphi}(0)$, and set $\tilde{\psi}^j{}'(0)$ and $\tilde{\varphi}'(0)$ to zero.
\item We integrate eq.\ \eqref{eq:SPNum} from $\tilde{r}=0$ to $\infty$.
\item We keep changing $\tilde{\varphi}(0)$ and repeating Step 2 until we find a smooth and nodeless $\tilde{\psi}^j(\tilde{r})$ solution.
\item The smooth and nodeless solution obtained in Step 3 has $\sum_j\tilde{M}^{\left(j\right)}=\lambda$. To normalize it, we rescale it according to eq.\ \eqref{eq:Resc}.
\end{enumerate}

The final solution from Step 4 is the soliton, see Fig.\ \ref{fig:Soliton}. Once we have it, we can extract $\tilde{E}$ with the help of the identity which follows from the Virial Theorem $\tilde{E}=-3\sum_j\int d^3\tilde{x} \tilde{\varphi}(\tilde{{\bf x}})\tilde{\psi^j}^2(\tilde{{\bf x}})$. Alternatively, a $(a\tilde{r}^{-1}-\tilde{E})$ fit to $\tilde{\varphi}(\tilde{r})$ at large distances yields the same result for $\tilde{E}$.

Using this prescription we found solutions which were indistinguishable from the ones in Appendix \ref{App:VDMGS}, see Fig.\ \ref{fig:Cartesian}, in terms of gravitational potential and energy density.

\begin{figure}[t] 
   \centering
   \includegraphics[width=3.45in]{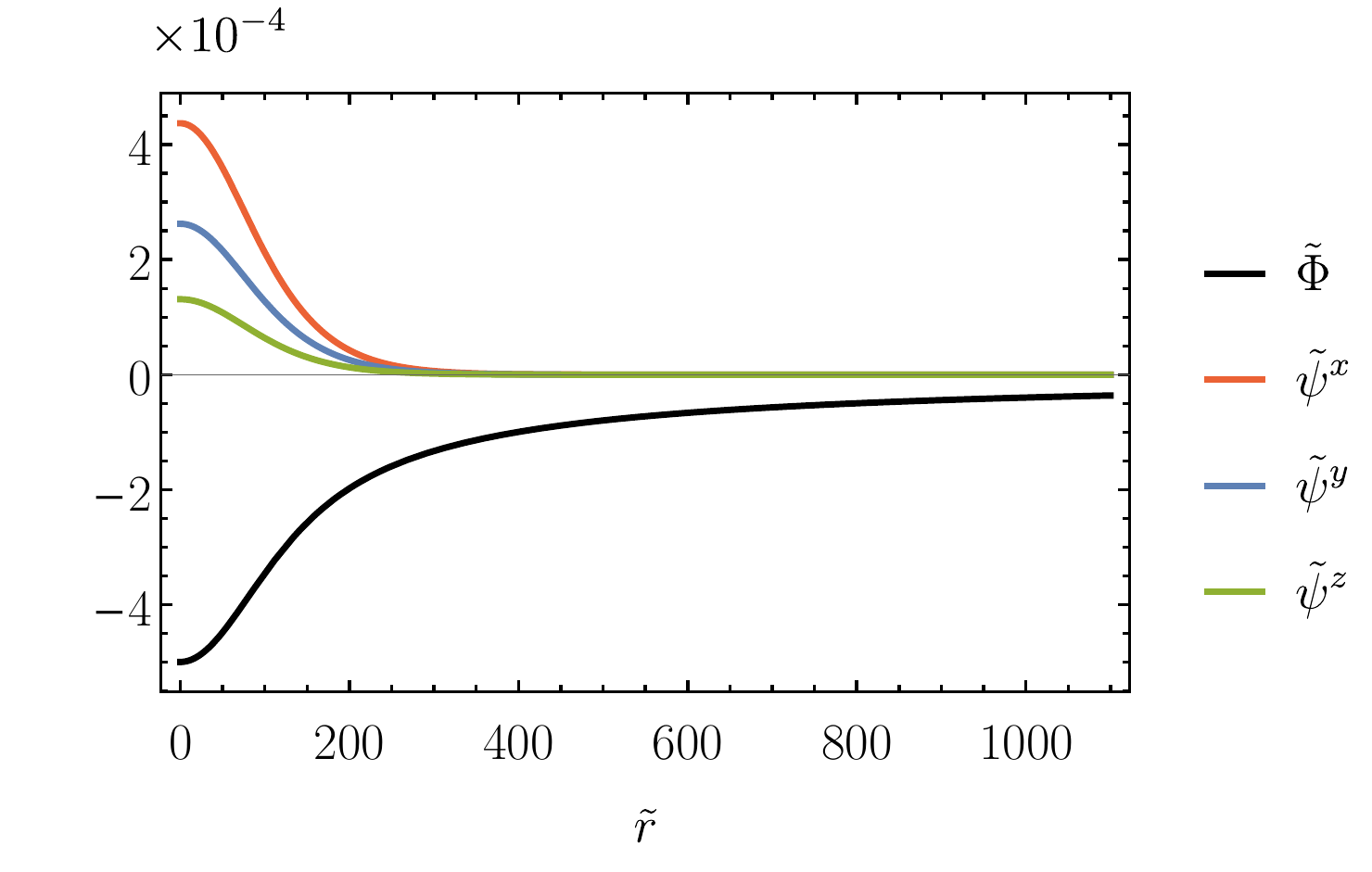}
   \caption{The ground-state solution for the ansatz in eq.\ \eqref{eq:CartesianDimLess1} with $\tilde{M}^{(x)}=0.690$, $\tilde{M}^{(y)}=0.248$. Note that $\sum_j{\tilde{M}}^{\left(j\right)}=1$. The $\tilde{\Phi}(\tilde{r})$ and $\sqrt{\sum_j\tilde{\psi}^{j2}(\tilde{r})}$ are identical to the blue and orange curves in Fig.\ \ref{fig:Soliton}, respectively. Since the data in Fig.\ \ref{fig:Soliton} is equivalent to, e.g., $\tilde{M}^{(x)}=1$, one concludes that the gravitational properties of the ground-state VDM halos, i.e., the gravitational potential, $\tilde{\Phi}(\tilde{r})$, and energy density, $\sum_j\tilde{\psi}^{j2}(\tilde{r})$, are independent of the $\tilde{M}^{\left(j\right)}$ values.}
   \label{fig:Cartesian}
\end{figure}

\section{Superfluid VDM, the Madelung equations and the virial theorem}
\label{App:SFVDM}

In this Appendix, for completeness, we provide details of the superfluid description and the virial theorem for VDM used above in Appendix \ref{App:Soliton}. For further details, we refer the reader to the original literature, see, e.g., references in \cite{Hui:2016ltb}.

\subsection{Superfluid description of VDM}
\label{App:SF}

The (super)fluid description of non-relativistic VDM is a generalization of the one of a non-relativistic scalar field \cite{Hui:2016ltb}. The three $\Aj$ fields have fluid densities and velocities defined as
\Beq
\label{eq:rhotheta}
\Aj\equiv\sqrt{\frac{\rho^{\left(j\right)}}{m}}e^{i\theta^{\left(j\right)}}\,,\qquad {\bf v}^{\left(j\right)}\equiv\frac{{\boldsymbol{\nabla}} \theta^{\left(j\right)}}{am}\,.
\Eeq
After substituting these definitions in the Schr\"{o}dinger equation for $\Aj$, eq.\ \eqref{eq:EOMSchrjFinal}, we arrive at the Madelung equations (for each component of $\Aj$)
\Beq
&\dot{\rho}^{(j)}+3H\rho^{\left(j\right)}+\frac{\boldsymbol{\nabla}}{a}\cdot\left(\rho^{\left(j\right)}{\bf v}^{\left(j\right)}\right)=0\,,\\
&\dot{\bf v}^{\left(j\right)}+H{\bf v}^{\left(j\right)}+\frac{1}{a}\left({\bf v}^{\left(j\right)}\cdot{\boldsymbol{\nabla}}\right){\bf v}^{\left(j\right)}\\
&\qquad\qquad\qquad=-\frac{{\boldsymbol{\nabla}}\Phi}{a}+\frac{1}{2a^3m^2}{\boldsymbol{\nabla}}\left(\frac{\nabla^2 \sqrt{\rho^{\left(j\right)}}}{\sqrt{\rho^{\left(j\right)}}}\right)\,.
\Eeq
The first equation is the FRW version of the continuity equation, whereas the second one represents the Euler equation of classical fluid dynamics, with the last term on the right hand side being the quantum pressure term.

For a self-gravitating vector dark matter, the superfluid version of the Poisson equation, eq.\ \eqref{eq:PoissonFinal}, is
\Beq
\label{eq:Poissonsuperfluid}
\frac{\Delta}{a^2} \Phi=\sum_j\frac{\rho^{\left(j\right)}}{2m_{\rm Pl}^2}\,.
\Eeq

\subsection{The Virial Theorem}\label{App:virial}

We now derive relations between the kinetic, gradient and gravitational energies for the three components of the superfluid VDM, which hold in stationary equilibrium (the relations are commonly known as the Virial Theorem). We consider a non-expanding universe. We follow the derivation for a single scalar FDM field from \cite{Hui:2016ltb} and extend it to VDM. 

Since $\rho^{\left(j\right)}=m^2|\Aj|^2$ is the mass density of the $j$th-component of the superfluid, the corresponding moment of inertia is given by
\Beq
I^{\left(j\right)}=\frac{1}{2}m\int d^3x |\Aj|^2x^2\,.
\Eeq
Using the Schr\"{o}dinger equation for $\Aj$, eq.\ \eqref{eq:EOMSchrjFinal}, with $a(t)=1$, one can derive the Virial Theorem
\Beq
\label{eq:VirialTheorem}
&\ddot{I}^{\left(j\right)}=\mathcal{V}^{\left(j\right)}+2\mathcal{K}^{\left(j\right)}+2Q^{\left(j\right)}\,,\\
&\mathcal{V}^{\left(j\right)}\equiv-\int d^3x\rho^{\left(j\right)} {\bf x}\cdot {\boldsymbol{\nabla}}\Phi\,,\\
&\mathcal{K}^{\left(j\right)}=\frac{1}{2}\int d^3x\rho^{\left(j\right)}|{\bf v}^{\left(j\right)}|^2\,,\\
&Q^{\left(j\right)}=\frac{1}{2m^2}\int d^3x|{\boldsymbol{\nabla}\sqrt{\rho^{(j)}}}|^2\,,
\Eeq
where $\mathcal{K}^{\left(j\right)}$ and $Q^{\left(j\right)}$ are interpreted as the kinetic and quantum energies of the $j$th-component of the superfluid.

In stationary equilibrium $\ddot{I}^{(j)}=0$, and since $\mathcal{K}^{\left(j\right)}\geq0$, we have
\Beq
\frac{Q^{\left(j\right)}}{|\mathcal{V}^{\left(j\right)}|}\leq\frac{1}{2}\,.
\Eeq
The bound is saturated if the phase of the wavefunction is position-independent,  ${\bf v}^{\left(j\right)}=\boldsymbol{\nabla}\theta^{\left(j\right)}=\boldsymbol{0}$. This is the case for the solitons discussed in Appendix \ref{App:Soliton}.

For a self-gravitating superfluid, the gravitational potential is given by
\Beq
\Phi({\bf x})=-\frac{1}{8\pi m_{\rm Pl}^2}\sum_j\int d^3y \frac{\rho^{\left(j\right)}({\bf y})}{|{\bf x}- {\bf y}|}\,.
\Eeq
Then, regardless of whether stationary equilibrium is established or not, there always exists a conserved total energy given by $W_{\rm tot}+\sum_j(\mathcal{K}^{\left(j\right)}+Q^{\left(j\right)})$, where the total gravitational potential energy of the self-gravitating superfluid is given by
\Beq
&W_{\rm tot}\equiv-\frac{1}{16m_{\rm Pl}^2}\int d^3x d^3y\frac{\rho_{\rm tot}({\bf x})\rho_{\rm tot}({\bf y})}{|{\bf x}- {\bf y}|}=\sum_j\mathcal{V}^{\left(j\right)}\,,\\
\Eeq
where
\Beq
\rho_{\rm tot}\equiv\sum_j\rho^{\left(j\right)}\,.
\Eeq

\bibliography{BibAuto,BibManual}

\end{document}